\documentclass[article]{revtex4}
\newcommand{\bi}{\begin{itemize}}
\newcommand{\ei}{\end{itemize}}
\newcommand{\be}{\begin{equation}}
\newcommand{\ee}{\end{equation}}
\newcommand{\ba}{\begin{eqnarray}}
\newcommand{\ea}{\end{eqnarray}}

\newcommand{\<}{\langle}
\renewcommand{\>}{\rangle}

\newcommand{\rhom}{\rho^{(m)}}
\newcommand{\rhor}{\rho^{(r)}}
\newcommand{\rhoim}{\rho_i^{(m)}}
\newcommand{\rhoir}{\rho_i^{(r)}}
\newcommand{\Ri}{^{(3)}{\cal{R}}_i}
\newcommand{\Riav}{\langle ^{(3)}{\cal{R}}_i\rangle}

\newcommand{\Dmi}{\Delta_i^{(m)}}
\newcommand{\Dri}{\Delta_i^{(r)}}
\newcommand{\Dki}{\Delta_i^{(k)}}

\begin{document}
\title{Adiabatic models of the cosmological radiative era.}
\author{Roberto A. Sussman$^\dagger$ and Mustapha Ishak$^\ddagger$}
\affiliation
{$^\dagger$
Instituto de Ciencias Nucleares,  Apartado Postal 70-543, UNAM, M\'exico DF,
04510, M\'exico.\\
$^\ddagger$ Department of Physics, Queen's University, Kingston, Ontario,
Canada,
K7L 3N6.}
\begin{abstract}
We consider a generalization of the Lemaitre-Tolman-Bondi (LTB) solutions
by keeping the LTB metric but replacing its dust matter source by an imperfect
fluid with anisotropic pressure $\Pi_{ab} $. Assuming that total matter-energy
density $\rho$ is the sum of a rest mass term, $\rhom$, plus a radiation
$\rhor=3p$ density where $p$ is the isotropic pressure, Einstein's equations
are fully integrated without having to place any previous assumption on the
form
of $\Pi_{ab} $. Three particular cases of interest are contained: the usual
LTB
dust solutions (the dust limit), a class of FLRW cosmologies (the
homogeneous limit)
and of the Vaydia solution (the vacuum limit). Initial conditions are provided
in terms of suitable averages and contrast functions of the initial
densities of $\rhom,\,\rhor$ and the 3-dimensional Ricci scalar along an
arbitrary initial surface $t=t_i$. We consider the source of the models as
an interactive
radiation-matter mixture in local thermal equilibrium that must be
consistent with
causal Extended Irreversible Thermodynamics (hence $\Pi_{ab} $ is shear
viscosity). Assuming
near equilibrium conditions associated with small initial density and
curvature contrasts,
the evolution of the models is qualitatively similar to that of adiabatic
perturbations on a
matter plus radiation FLRW background. We show that initial conditions
exist that lead to
thermodynamically consistent models, but only for the full transport
equation of Extended
Irreversible Thermodynamics. These interactive mixtures provide a
reasonable approximation to a
dissipative 'tight coupling' characteristic of radiation-matter mixtures in
the radiative
pre-decoupling era.
\end{abstract}
\maketitle
\section {Introduction}
The Lemaitre-Tolman-Bondi (LTB) solutions with a dust source
\cite{ksmh},\cite{kras},
\cite{ellisvan} are widely popular models of cosmological inhomogeneities (see
\cite{kras} for a comprehensive review). We present a
generalization in which the LTB metric is kept but the source is replaced by an
imperfect fluid with anisotropic pressure, under the assumption that
matter-energy
density is decomposed as
$\rho=\rhom+\rhor$, a mixture of ``rest mass'' and ``radiation''
components (a mixture of non-relativistic and relativistic matter), so that
$\rhor=3p$, where $p$ is the isotropic pressure.
The purpose of this paper is to derive the important geometric
properties of the solutions within a  convenient framework and then to
examine the compatibility of the models with the physics of
radiation-matter sources.  

The study of inhomogeneous cosmological models such the LTB models
is a well motivated and justified endeavor. First of all, it complements the
usual perturbative approach by allowing one to study the non-linear
evolution of inhomogeneities. Also, a nearly isotropic Cosmic
Microwave Background Radiation (CMB)  does not rule out
an inhomogeneous universe compatible with current CMB observations
\cite{clarbar},\cite{MHE}. Moreover, the models we examine here are of
particular interest when it is necessary
to consider the radiation component and the dissipative processes
associated with its interaction with non-relativistic matter 
\cite{pad}, \cite{peac}, \cite{bor}.
On the other hand, even if it is reasonable to use a dust source model as a
theoretical matter model for present day universe, this source cannot
describe the CMB and does not
allow one to deal with temperatures of the photon gas.

Various physical interpretations and observational bounds have been
proposed for
anisotropic stresses in cosmic matter sources\cite{barrow},\cite{barmaa}.
If the
mixture components are not interacting, then we have a mixture of collision-less
non-relativistic matter and CMB with
anisotropic
pressure understood as the quadrupole moment of the distribution function
for the
photon gas\cite{MES1} - \cite{chal_las} under a Kinetic Theory
approach\cite{EMT},\cite{ETM}. For a matter-radiation mixture in which there is
interaction between the components (a radiative photon-electron
interaction), the
anisotropic pressure can be interpreted as shear viscosity within a causal
irreversible thermodynamic approach \cite{sw},
\cite{wei} -\cite{JCL}. This provides an adequate model for the radiative
era of
cosmic evolution, from after nucleosynthesis to up to ``matter'' and
``radiation''
decoupling. 

In section II, Einstein's field equations are integrated up to a
Friedmann-like equation, a quadrature in which the free parameters are
three initial
value functions related to the average of initial energy densities
$\rhoim,\,\rhoir$ and $^{(3)}R $, the 3-dimensional Ricci scalar along
the initial surface $t=t_i$. 
In section III, we describe the possible 
physical interpretations of the models source focusing on the interacting
matter-radiation mixture. In sections IV and V we express the free
parameters in terms of suitable volume averages and initial contrast
functions defined along $t=t_i$, leading to simplified and elegant forms
for all the relevant geometric and physical quantities. These quantities
become fully determined (up to initial conditions) once the Friedmann-like
quadrature is integrated (appendix A) yielding canonical and  parametric
solutions. The models contain three important particular cases
presented in appendix B.
In sections VI and VII, we examine the state variables of the model
under small density contrasts. Theses small contrasts lead to two important
classes of initial conditions equivalent to the definition of
adiabatic and quasi-adiabatic perturbations of the initial densities
and the initial curvature. In section VIII, we study the
thermodynamic consistency of the coupled mixture model. 

Different variants of the imperfect fluid generalization of LTB solutions have
been presented previously: using the equation of state of a non-relativistic
ideal gas \cite{suss}, considering various possible ideal gas equations of
state
and a generalization to non-spherical geometry of the Szekeres-Szafron type
\cite{susstrig} and the parabolic case for the matter-radiation mixture
\cite{mixt}. The present paper extend and complement the
results of previous literature with respect to the specific cases of
spherically symmetric and curved (elliptic and hyperbolic) models with a
radiation-matter mixture source. 
\section{Field equations.}
Consider the usual LTB metric
\be
 \hbox{d}s^2= -c^2\hbox{d}t^2+ \frac{Y'^{2}}{1-K} \,\hbox{d}r^2
+Y^2\left[\hbox{d}\theta^2+\sin^2 (\theta) \hbox{d}\phi^2
\right],\label{gab}
\ee
where $Y = Y(t,r)$,\, $K=K(r)$ and a prime denotes derivative with respect
to $r$.  Instead
of the usual dust source, we shall consider the stress-energy tensor of a
fluid with anisotropic
pressure
\be
T^{ab}= \rho u^au^b+ph^{ab}+\Pi^{ab}, \quad h^{ab}= c^{-2}u^a
u^b+g^{ab},\quad \Pi^a\,_a= 0 \, ,
\label{Tab}
\ee
where the  most general form for the anisotropic pressure tensor for
the metric (\ref{gab}) with matter source (\ref{Tab}) is given by:
$\Pi^a\,_b={\bf{\hbox{diag}}}\left[0,-2P,P,P \right]$, with $P = P(t,r)$ to
be determined by the
field equations.  Einstein's field equations for (\ref{gab}) and
(\ref{Tab}) are
\be\frac{8\pi G}{c^4} \rho \ = \ -G^t\,_t \ = \ \frac{\left[Y(\dot
Y^2+K)\right]'}{Y^2Y'},\label{eqrho1}
\ee
\be\frac{8\pi G}{c^4} p \ = \ \frac{1}{3}(2G^\theta\,_\theta+G^r\,_r) \ = \
-\frac{\left[Y(\dot
Y^2+K)+2Y^2\ddot Y\right]'}{3Y^2Y'},\label{eqp1}
\ee
\be\frac{8\pi G}{c^4} P \ = \
\frac{1}{3}(G^\theta\,_\theta-G^r\,_r) \ = \ -\frac{Y}{6Y'}\left[\frac{Y(\dot
Y^2+K)+2Y^2\ddot Y}{Y^3}\right]',\label{eqP1}\ee
where $\dot{Y}=u^aY_{,a}=Y_{,ct}$. In order to integrate these equations we
need to impose a
relation between $\rho$ and $p$ but no previous assumption on $P$ is necessary.

Consider $\rho$ in (\ref{Tab}) to be the sum of two contributions:
non-relativistic
matter described by dust plus radiation energy density ($\rhor$)
\be\rho \ = \ \rhom \ + \ \rhor,\quad
p \ = \ \frac{1}{3}\rhor,\quad\quad\hbox{with:}\quad
\frac{8\pi G}{c^4}\,\rhom \ = \ \frac{2M'}{Y^2Y'},\label{eqrhop}\ee
where $M=M(r)$, while the form of $\rhor$ will be discussed further
ahead. Inserting (\ref{eqrhop}) into (\ref{eqrho1}) and (\ref{eqp1}) and
integrating once with
respect to $t$ leads to the Friedmann equation
\be\dot Y^2 \ = \ \frac{1}{Y}\left[\,2M\,+\,W\frac{Y_i}{Y}\,\right] \ - \
K,\label{Ydot} \ee
where $W=W(r)$ and $Y_i=Y(t_i,r)$ for an arbitrary fixed value $t=t_i$. The
interpretation for
$M$ and $W$ follows by substituting (\ref{Ydot}) into equations
(\ref{eqrho1}), (\ref{eqp1})
and (\ref{eqP1})
\ba
\frac{8\pi G}{c^4}\,\rho \,Y^2Y' \ = \
\left[\,2M+W\frac{Y_i}{Y}\,\right]',\nonumber \\  
\frac{8\pi G}{c^4}\,p \,Y^2Y' \ = \
\frac{1}{3} \left[\,W\frac{Y_i}{Y}\,\right]',\nonumber \\ 
\frac{8\pi G}{c^4}\,P\frac{Y'}{Y} \ = \ -
\frac{1}{6}\left[\,\frac{WY_i}{Y^4}\,\right]'.\label{eqrhopP}\ea
Since $Y^2Y'$ is proportional to the determinant of the spatial part of the
metric (\ref{gab}),
it is a covariant measure of proper local volumes. Therefore $M'$ and $W'$
must have units of
length and so it is convenient to define them in terms of the initial
energy densities of
matter and radiation
\be
2M \ = \ \frac{8\pi G}{c^4}\,\int{\rhoim Y_i^2Y_i'dr},
\quad W \ = \ \frac{8\pi G}{c^4}\int{\rhoir Y_i^2Y_i'dr},\quad
\rhoir \ = \ 3p_i,\label{eqMW}
\ee
where $\rhoim,\,\rhoir$ are $\rhom,\,\rhor$ evaluated at
$t=t_i$.

For a non-rotating fluid with a geodesic 4-velocity the two
nonzero kinematic parameters are: the expansion scalar, $\Theta\equiv
u^a\,_{;a}$, and the shear
tensor, $\sigma_{ab}= u_{(a;b)}-(\Theta/3)h_{ab}$. These parameters for
(\ref{gab}) take the
form
\be \Theta \ = \ \frac{2\dot Y}{Y}\,+\,\frac{\dot Y'}{Y'},\label{eqTheta1}\ee
\be\sigma^a\,_b \ = \ \hbox{diag}\,[0,-2\sigma,\sigma,\sigma], \quad
\sigma \ \equiv
\ \frac{1}{3}\left(\frac{\dot Y}{Y}-\frac{\dot Y'}{Y'}\right),\quad
\sigma_{ab}\sigma^{ab} \ = \ 6\sigma^2,\label{eqsigma1}\ee
Under the assumptions (\ref{eqrhop}) the energy and momentum balance law,
$T^{ab}\,_{;b}= 0$,
associated with (\ref{gab}) and (\ref{Tab}) gives
\ba \dot \rhor \ + \ \frac{4}{3}\,\Theta\,\rhor \ + \ 6\,\sigma\,P \ = \
0,\nonumber\ea
\ba h_a^b\,(\,p_{,b}+\Pi^c\,_{b;c}\,) \ = \ 0 \quad \Rightarrow\quad
(\,p-2P\,)'-6P\frac{Y'}{Y} \ = \ 0,\nonumber\ea
clearly illustrating how the pressure gradient is exactly balanced by the
divergence of $\Pi^a\,_b$, allowing for non-zero pressure gradients to be
compatible with geodesic motion of comoving matter.
\section{Matter-radiation models.}
The relation between $\rho$ and $p$ given by (\ref{eqrhop}) suggests that
the matter
source can be understood as a mixture of non-relativistic matter (to be
referred as
``matter'') and ultra-relativistic matter (to be referred as
``radiation''), both characterized
by the same 4-velocity. Therefore, a suitable physical interpretation for
this source is
an interactive, `tightly coupled' mixture of matter and radiation examined
under a
hydrodynamical approach. The anisotropic pressure becomes a dissipative
term (a shear viscosity)
that must be examined within a thermodynamic framework\cite{wei} -\cite{JCL},
\cite{IS},\cite{DDC}. This type of source provides a convenient description
for the ``radiative
era'', after nucleosynthesis and before decoupling of radiation and matter.
An alternative
approach is that of a decoupled mixture (non-relativistic matter plus the
CMB) under the
framework of Kinetic Theory \cite{MES1} - \cite{ETM}. In this paper we
shall consider only the
hydrodynamical approach applicable to the radiative era, leaving the study
of a decoupled mixture
for a future work.

The `tightly coupled' mixture of non-relativistic matter 
and radiation (a photon gas) is characterized by local thermal equilibrium
(common temperature)
among the components. This situation implies that interaction timescales
are smaller than the
cosmic expansion timescale, thus a hydrodynamical approach is valid so that
this
interactive mixture behaves as a single dissipative fluid. Since heat flux
necessarily vanishes
for the LTB metric (\ref{gab}), this fluid can be described by the
momentum-energy tensor
(\ref{Tab}) where $\Pi^{ab}$ now becomes the shear viscous tensor.
Considering that
non-relativistic matter can be modeled by a classical monatomic ideal gas,
while the photon gas
satisfies the Stefan-Boltzmann law, the tightly coupled mixture of these
components requires
that $\rho$ and $p$ in (\ref{Tab}) must comply with the equation of state
\be \rho \ = \ m\,c^2\,n \ + \ \frac{3}{2}\,n\,k\,T\,+\,a\,T^4,\quad p \ =
\ n\,k\,T\,+\,\frac{1}{3}\,a\,T^4,\label{eqst1}\ee
where $k,\,a$ are Boltzmann and Stephan-Boltzmann constants, $T$ is the common
mixture temperature, $m$ is the mass of the most representative
species of non-relativistic particles and $n$ is particle number density,
satisfying (we assume there is no net creation or annihilation of
particles) the
conservation law
\be(n\,u^a)_{;a} \ = \ 0\quad\Rightarrow \quad n \ = \
\frac{N(r)}{Y^2Y'},\label{eqmatcons}\ee
where $N(r)$ is an arbitrary function. If $nkT\ll aT^4$, or equivalently
$aT^3/nk\gg
1$, but $mc^2n/aT^4$ is not negligible, then (12) can be approximated
by\cite{sw},\cite{wei}
\be \rho \ = \ m\,c^2\,n\,+\,a\,T^4,\quad p \ = \
\frac{1}{3}\,a\,T^4,\label{eqst2}\ee
Comparing the conservation
law (\ref{eqmatcons}) and the equation of state (\ref{eqst2}) with (6a) and
(6b), it is evident
that the generalized LTB solutions can provide a model for the
radiation-matter tightly
coupled mixture if  we identify
\be \rhom \ = \ m\,c^2\,n,\quad m\,c^2\,N \ = \ 2\,M' \ = \
\frac{8\pi G}{c^4}\,\rhoim\,Y_i^2Y_i',\quad
\rhoim\ = \ m\,c^2\,n_i,\label{eqrhom1}\ee
so that the radiation energy density and the temperature are given by
\be \frac{\left[W\,Y_i/Y\right]'}{Y^2Y'} \ = \ \frac{8\pi G}{c^4}\,\rhor \
= \ \frac{8\pi
G}{c^4}\,a\,T^4,\quad \frac{8\pi G}{c^4}\,a\,T_i^4 \ = \
W'\,Y_i^2Y_i'.\label{eqrhor1}\ee
As an alternative to (\ref{eqst1}) and (\ref{eqst2}), it is possible to
consider, instead of the
Stefan-Boltzmann law, $\rhor=aT^4$, the energy density and pressure of the
radiation component as those of an ideal ultra-relativistic gas
\be p^{(r)} \ = \ n^{(r)}\,k\,T,\quad \rhor \ = \ 3p^{(r)},\quad n^{(r)} \
= \ \frac{N^{(r)}}{Y^2Y'}\label{eqstig}\ee
where $n^{(r)}$ is the corresponding particle number density, independently
satisfying a
conservation law like (\ref{eqmatcons}) with $N^{(r)}= N^{(r)}(r)$. This
approach has been
followed previously in \cite{susstrig} and \cite{mixt}, the advantage being
a simpler
expression than (\ref{eqrhor1}) for the temperature in terms of the
gradients of $Y$
\be \frac{8\pi G}{c^4}\,p\,Y^2Y' \ = \
\frac{1}{3}\,\left[\,W\frac{Y_i}{Y}\,\right]' \ = \
N^{(r)}\,k\,T,\label{eqTig1}\ee
In global thermal equilibrium, the fact that photon entropy per barion
is conserved: $s^{(e)}=(4/3)\,a\,T^3/n=\hbox{const.}$ implies that the ratio
of photons to baryons is constant and $a\,T^4\propto n\,k\,T$, hence the
Stefan-Boltzmann and
ideal gas laws are equivalent. Since we shall consider near equilibrium
conditions in which
these two laws should be almost (but not exactly) equivalent, then
$a\,T^3/(n\,k)$ is
proportional to the (approximately constant) number of photons per
non-relativistic particle. If
the latter are baryons and electrons, then this quantity is very large,
thus justifying the
approximation leading from (\ref{eqst1}) to (\ref{eqst2}). If temperatures
are high enough so
that the ratio $aT^4/(mc^2n)$ is low enough for creation/annihilation
processes to cancel each
other (hence (\ref{eqmatcons}) holds), then (\ref{eqst2}) provides a
reasonable description of
cosmic matter in the radiative era. The dominant radiative
processes characteristic of
this era mostly involve the photon-electron interaction (Thomson and
Compton scattering,
Brehmstrallung, etc) \cite{pad}, \cite{wei} and \cite{JCL}. The
temperature range for the
radiative era (up to matter-radiation decoupling) is roughly $4\times 10^3
K. <T<10^6 K.$
The anisotropic pressure $\Pi^a\,_b$ is now the shear viscosity tensor,
hence its evolution law must be consistent with the shear viscosity
transport equation
of a causal irreversible thermodynamic theory. In the particular case
when shear
viscosity is the single dissipative flux, the entropy per particle and shear
viscosity transport equation provided by Extended Irreversible
Thermodynamics are
\be s \ = \ s^{(e)}\,+\,\frac{\alpha}{nT}\,\Pi_{ab}\Pi^{ab},\quad
\Rightarrow \quad
(snu^a)_{;a} \ = \ \dot s nu^a \ \geq \ 0,\label{eqs1}\ee
\be \tau \,\dot\Pi_{cd}\,h^c_ah^d_b\,+\,\Pi_{ab}\left[
1+{1\over2}T\eta\left({{\tau}\over{T\eta}}  \,
u^c\right)_{;c}\right]\,+\, 2\,\eta\,\sigma_{ab} \ = \ 0,\label{eqtr1}
\ee
where $s^{(e)}=(4/3)aT^3/n $ is the equilibrium entropy per particle, taken
(approximately) as the
initial photon entropy per non-relativistic particle, $\eta,\,\tau$ are the
coefficient of shear
viscosity and a relaxation time which, together with $\alpha$, are
phenomenological coefficients
whose functional form depends on the fluid under consideration. A
``truncated'' version of
(\ref{eqtr1}), also known as the ``Maxwell-Cattaneo'' transport equation,
is often
used\cite{JCL}, \cite{IS} for the sake of mathematical simplicity:
\be \tau \,\dot\Pi_{cd}\,h^c_ah^d_b\,+\,\Pi_{ab}\,+\, 2\,\eta\,\sigma_{ab}
\ = \ 0,\label{eqtrmc1}
\ee
Although this truncated equation satisfies causality and stability
requirements, numerical and theoretical studies indicate that it might be
problematic\cite{truncated}. In spite of the obvious limitation of having
only one dissipative
flux, the generalized LTB models provide arguments to infer the ranges of
applicability of the
truncated and full equations.

The application of the thermodynamic formalism to the generalized
LTB models, considered as a models of a hydrodynamical `tightly coupled'
mixture of matter and
radiation, requires a convenient selection of the phenomenological quantities
$\eta,\,\alpha,\,\tau$, and then solving the appropriate transport equation
(either
(\ref{eqtr1}) or (\ref{eqtrmc1})). In order to proceed with this task we
will find in the following sections an intuitive characterization of
initial conditions and suitable forms for the state variables.
\section{Initial conditions and volume averages}
It is convenient to transform the Friedmann equation (\ref{Ydot}) into the
simpler
quadrature
\be\dot y^2 \ = \ \frac{-\kappa\,y^2+2\mu \,y +\omega}{y^2},\label{ydot} \ee
where
\ba y \ = \ \frac{Y}{Y_i},\quad \mu \ \equiv \ \frac{M}{Y_i^3},\quad
\omega \ \equiv \ \frac{ W}{Y_i^3},\quad \kappa \ \equiv \
\frac{K}{Y_i^2}, \quad Y_i \
= \ Y(t_i,r)\nonumber
\ea
Bearing in mind (\ref{eqMW}), assuming that the initial hypersurface
$t=t_i$ is everywhere regular and contains at least a symmetry center
\cite{mixt} \cite{susstrig}, we can express the quantities $\mu,\,\omega,\,\kappa$ 
as the following type of volume averages along $t=t_i$
\ba  \< \rhoim\>  \ \equiv \ \frac{\int{\rhoim
Y_i^2Y_i'dr}}{\int{Y_i^2Y_i'dr}}
\quad \Rightarrow \quad 2\,\mu \ = \ \frac{8\pi G}{3c^4}\,\<
\rhoim\>,
\cr
\cr
\cr
\< \rhoir\> \ \equiv \ \frac{\int{\rhoir Y_i^2Y_i'dr}}{\int{Y_i^2Y_i'dr}}
\quad \Rightarrow \quad \omega \ = \ \frac{8\pi G}{3c^4}\,\< \rhoir\>,
\cr
\cr
\cr
\Riav \ \equiv \ \frac{\int{\Ri Y_i^2Y_i'dr}}{\int{Y_i^2Y_i'dr}}
\quad \Rightarrow \quad \kappa \ = \
\frac{1}{6}\,\Riav,\label{<i>},\label{volave}\ea
where
\ba \Ri \ = \ \frac{2(K\,Y_i)'}{Y_i^2Y_i'},\nonumber\ea
is the Ricci scalar of (1) evaluated at $t=t_i$ and the range of
integration goes from the
symmetry center up to a comoving sphere marked by $r$.
Notice that these
averages (save for $K= 0$) do not coincide with the local proper volume
defined by
(\ref{gab}), namely: $Y_i^2Y_i'/\sqrt{1-K}=(1/3)d(Y_i^3)/\sqrt{1-K}$,
though they can also be
characterized covariantly, since the metric function $Y$ (and so $Y_i$) is,
in spherical
symmetry, the ``curvature radius'', or the proper radius of the orbits of
the rotation group
SO(3). The volume averages (\ref{volave}) lead to a compact and elegant
mathematical description of initial conditions.
From the volume averages in (\ref{volave}) we can define the 
functions $\Dmi,\,\Dri,\,\Dki$ satisfying the following appealing
relations
\ba \rhoim=\<\rhoim\>\left[1+\Dmi\right], \nonumber \\
\rhoir=\<\rhoir\>\left[1+\Dri\right], \nonumber \\
\Ri= \Riav\left[1+\Dki\right], \label{eqDeltas2}
\ea
justifying their interpretation as initial density and curvature
``contrast functions'', as they provide a measure of the contrast of
initial value functions
$\rhoim,\,\rhoir,\,\Ri$ with respect to their volume averages
along the initial hypersurface $t=t_i$. Because of their definition in
terms of the volume averages (\ref{<i>}), the value of the contrast
functions at a given $r=r_*$ depend on the values of the functions
$\rhoim,\,\rhoir,\,\Ri$ on the integration range
$0\leq r\leq r_*$, in which we assume that $r=0$ marks a symmetry center
characterized by $Y(t,0)=\dot Y(t,0)=0$ \cite{mixt} \cite{susstrig}. Therefore, for a rest
mass density lump $\rhoim$ decreases with increasing $r$, while for a rest
mass density void it increases with increasing $r$, hence (from
(\ref{eqDeltas2})) we have
$-1\leq \Dmi\leq 0$ for a lump and $\Dmi\geq 0$ for a void. The
same criterion holds for the initial radiation density and scalar curvature,
$\rhoir$ and $\Ri$.

From (\ref{<i>}) and (\ref{eqDeltas2}), it is straightforward to verify
that the functions
$\rhoim,\rhoir,\Ri$ relate to $\Dmi,\,\Dri,\,\Dki $ by
\be \epsilon_1 \ = \ \frac{\omega}{\mu} \ = \
\frac{2\rhoir}{\rhoim}\,\frac{1+\Dmi}{1+\Dri},\quad \epsilon_2 \ = \
\frac{\kappa}{\mu} \ = \
\frac{c^4\,\Ri}{8\pi G\,\rhoim}\,\frac{1+\Dmi}{1+\Dki},\label{eqeps}\ee
and
\be \Delta_i^{(r)} \ = \ \frac{\omega'/\omega}{Y'_i/Y_i}, \quad
\Delta_i^{(m)} \ =
\ \frac{\mu'/\mu}{Y'_i/Y_i},\quad \Delta_i^{(k)} \ = \
\frac{\kappa'/\kappa}{Y'_i/Y_i}.\label{eqDeltas1}\ee
Since the state variables $\rhom,\,\rhor,\,P$ and kinematic parameters
$\Theta,\sigma$ are given in terms of quantities that will depend on the initial value
functions $\mu,\,\omega,\,\kappa$ and their gradients, it is useful to be able to express
these gradients, by means of (\ref{eqeps}) and (\ref{eqDeltas1}), in terms of the more
intuitively appealing  initial contrast functions defined by (\ref{eqDeltas2}). We examine
this point in the following section. 
\section{State variables and kinematic parameters}
Using (\ref{eqrhopP}) (\ref{eqrhom1}) and (\ref{eqrhor1}), we can rewrite
the state variables $\rhom,\,\rhor,\,P$ in a more appealing form
\be
\rhom \ = \ m\,c^2\,n \ = \ \frac{\rhoim}{y^3\Gamma}, \quad \rhoim \ = \
m\,c^2\,n_i
\label{eqrhop2}\ee
\ba \rhor \ = \ 3\,p \ = \ \frac{\rhoir\Psi}{y^4\Gamma},\quad \rhoir \ = \
a\,T_i^4\quad \hbox{(Stefan-Boltzmann)},\nonumber \\ 
\rhoir \ = \ 3n_i^{(r)}\,k\,T_i\quad \hbox{(Ideal Gas)},\ea
\be T \ = \ \frac{T_i}{y} \left(\frac{\Psi}{\Gamma}\right)^{1/4} \
\hbox{(Stefan-Boltzmann)},\ T \ = \ \frac{T_i\,\Psi}{y} \
\hbox{(Ideal Gas)},\label{eqT}\ee
\be P \ = \ \frac{\rhoir\Phi}{6y^4\Gamma},\label{eqP2} \ee
where the auxiliary functions $ \Gamma,\,\Psi,\,\Phi$,  characterizing the
spacial dependence of $\rho,\,p,\,P$, are given by
\be\Gamma=\frac{Y'/Y}{Y_i'/Y_i}=1+\frac{y'/y}{Y_i'/Y_i},\quad
\Psi=\frac{4+3\Delta_i^{(r)}-\Gamma}{3[1+\Delta_i^{(r)}]},\quad
\Phi=\frac{4+3\Delta_i^{(r)}-4\Gamma}{3[1+\Delta_i^{(r)}]},\label{eqGPP}\ee
where $\Dri$ has been defined in (\ref{eqDeltas2}).
The function $\Gamma$ can be obtained as a function of
$y,\,\mu,\,\omega,\,\kappa$ and the initial contrast functions by
performing the integral quadrature of (\ref{ydot}). Let
\be Z\ \equiv \ \int{\frac{ydy}{\left[-\kappa y^2+2\mu
y+\omega\right]^{1/2}}} \ =
\ \frac{1}{\sqrt{\mu}}\,\int{\frac{ydy}{\left[-\epsilon_2\,y^2+
2\,y+\epsilon_1\right]^{1/2}}},\label{eqZ}\ee
where
\ba Z_i(\mu,\omega,f) \ = \ Z|_{y=1},\quad \epsilon_1 \ \equiv \
\frac{\omega}{\mu},
\quad \epsilon_2 \ \equiv \ \frac{\kappa}{\mu}, \nonumber \ea
then, by using the chain rule on $Z-Z_i$ we can obtain for $\Gamma$, $\Psi$
and $\Phi$ in (\ref{eqGPP}) the following expressions
\be \Gamma \ = \ 1 \ - \ 3\,A\,\Delta_i^{(m)} \
- \ 3\,B\,\Delta_i^{(r)} \ - \ 3\,C\,\Delta_i^{(k)},\label{eqGamma1}
\ee
\be \Psi \ = \ \frac{1 \ + \ A\,\Delta_i^{(m)} \ + \ (1+B)\,\Delta_i^{(r)}
\ + \
C\,\Delta_i^{(k)}} {1 \ + \ \Delta_i^{(r)}},\label{eqPsi1}\ee
\be \Phi \ = \ \frac{4\,A\,\Delta_i^{(m)} \ + \ (1+4B)\,\Delta_i^{(r)} \ +
\ 4\,C\,\Delta_i^{(k)}}
{1 \ + \ \Delta_i^{(r)}},\label{eqPhi1}\ee
where
\be A  =  \frac{\mu\,[\partial(Z-Z_i)/\partial\mu]}{y\,[\partial
Z/\partial y]},\
B  =  \frac{\omega\,[\partial(Z-Z_i)/\partial\omega]}{y\,[\partial
Z/\partial y]},\
C  =  \frac{\kappa\,[\partial(Z-Z_i)/\partial\kappa]}{y\,[\partial
Z/\partial y]},
\label{eqABC}\ee
In order to obtain the explicit functional forms for $A,B,C$ above we need
to evaluate $Z$
explicitly. This is done in Appendix A, while a convenient interpretation
as ``initial contrast functions'' has been 
provided in the previous section for the functions $\Delta_i^{(m)}, \, \Delta_i^{(r)},
\, \Delta_i^{(k)}$.
It is important to mention that no {\it a priori} assumption on $P$ was
made in order to obtain
(\ref{eqP2}). This specific form of anisotropic pressure (shear viscosity)
follows directly from
(\ref{eqrho1}), (\ref{eqp1}), (\ref{eqst2}) and (\ref{ydot}).

In terms of the new variables, the kinematic parameters, $\Theta,\,\sigma$,
introduced in (\ref{eqTheta1}) and (\ref{eqsigma1}), take the form
\ba \Theta  & = &  \frac{3\dot y}{y} \ + \ 
\frac{\dot\Gamma}{\Gamma} \nonumber \\
            & = &  \frac{\sqrt{Q}}{y}\, \frac{1-(3A+yA_{,y})\Delta_i^{(m)}-
(3B+yB_{,y})\Delta_i^{(r)}-(3C+yC_{,y}) \Delta_i^{(k)}}
{1-3A\Delta_i^{(m)}-3B\Delta_i^{(r)}-3C\Delta_i^{(k)}}\label{eqTheta2}\ea
\be \sigma \ = \ -\frac{\dot\Gamma}{3\Gamma} \ = \ \frac{\sqrt{Q}}{y}\,\,
\frac{A_{,y}\Delta_i^{(m)}+B_{,y}\Delta_i^{(r)}+C_{,y}\Delta_i^{(k)}}
{1-3A\Delta_i^{(m)}-3B\Delta_i^{(r)}-3C\Delta_i^{(k)}},\label{eqsigma2}\ee
where $Q\equiv -\kappa y^2+2\mu y +\omega$ and a sub-index $_y$ means partial
derivative with
respect to $y$.  The new solutions presented so far become fully determined
once the Friedmann
equation (\ref{ydot}) is integrated ($Z$ is explicitly known) for specific
initial conditions
provided by $\mu,\,\omega$ and $\kappa$. This integration is presented in
appendix A and particular cases of interest are presented in appendix
B. The conditions for the models to comply with regularity at
the center and energy conditions are given in \cite{mixt} and
\cite{susstrig}.
\section{Small density contrasts and an adiabatic evolution.}
The thermodynamic study of the generalized LTB models, under the
framework of Extended
Irreversible Thermodynamics, necessarily assumes small deviations from
equilibrium. Since the
equilibrium limit of the models (excluding the dust limit which has trivial
thermodynamics) is
the FLRW limit that follows by setting the initial contrast functions to
zero, then a ``near
equilibrium'' evolution can be related to a ``near FLRW'' evolution that
follows by assuming
``near homogeneous initial conditions'', or in other words, small initial
contrast functions,
{\it ie}:
\be |\Dmi| \ \ll \ 1, \quad |\Dri| \ \ll \ 1, \quad |\Dki| \ \ll \
1,\label{small_D1}\ee
implying
\ba \rhoim \ \approx \ \<\rhoim\>,\quad
\rhoir \ \approx \ \<\rhoir\>,\quad \Ri \ \approx
\ \<\Ri\>, \cr\cr\cr2\mu \ \approx \ \frac{8\pi G}{3c^4}\,\rhoim,\quad
\omega \ \approx
\ \frac{8\pi G}{3c^4}\,\rhoir,\quad \kappa \ \approx \
\frac{1}{6}\,\Ri,  \cr\cr\cr \epsilon_1 \ \approx \
\frac{2\rhoir}{\rhoim}, \quad \epsilon_2 \ \approx \
\frac{c^4}{8\pi G}\,\frac{\Ri}{\rhoim}, \quad\quad (\epsilon_1,\epsilon_2\
\approx \ \hbox{const.}),\label{small_D2}\ea
where the exact form of $ \epsilon_1,\epsilon_2$ is given by (\ref{eqeps}).
Under these
assumptions, all functions depending on $(\mu,\omega,\kappa,y)$ become
approximately functions of
$y$ only. From (\ref{eqZ}), (\ref{eqGamma1}), (\ref{eqPsi1}),
(\ref{eqPhi1}) and (\ref{eqABC}),
this implies
\be Z \approx Z(y), \  t-t_i \approx Z(y)-Z_i,\ A \approx A(y), \
B \approx B(y), \ C \approx C(y), \label{small_D3}\ee
with $Z_i\approx \hbox{const.}$, so that hypersurfaces $y=\hbox{const.}$
approximate
hypersurfaces $t=\hbox{const.}$ and $y$ becomes the time parameter. The
radial dependence in
quantities like $\Gamma,\,\Psi,\,\Phi$ is then contained in the initial
contrast functions. Regarding the state variables, a Taylor series
expansion around
$\Dmi,\,\Dri,\,\Dki$ yields at first order
\be \frac{1}{\Gamma} \ \approx 1 + \delta, \quad
\frac{\Psi}{\Gamma} \ \approx \ 1+\frac{4}{3}\,\delta, \quad
\frac{\Phi}{\Gamma} \ \approx \
\Delta_i^{(r)}+\frac{4}{3}\,\delta,\label{small_D_pert1}\ee
where we can formally identify the ``perturbation'' as
\be  \quad \delta \ \equiv \
3\,[A\,\Dmi+B\,\Dri+C\,\Dki] \ = \ 1-\Gamma, \label{eqdelta}\ee
so that, from (\ref{eqrhop2}) and (\ref{small_D_pert1}), we have
\be \rhom  \approx  \frac{\rhoim}{y^3}\,\left[\,1 + \delta\,\right],\
\rhor  \approx  \frac{\rhoir}{y^4}\,\left[\,1 +
\frac{4}{3}\,\delta\,\right], \  P  \approx 
\frac{\rhoir}{6y^4}\,\left[\,\Dri +
\frac{4}{3}\,\delta\,\right],\label{small_D_pert2}\ee
Therefore, under the ``small contrast approximation'' given by
(\ref{small_D1}),
(\ref{small_D2}) and (\ref{small_D3}) the models follow a nearly
homogeneous evolution with
respect to a scale factor $y\approx y(t)$, deviating from homogeneity in a
way (as given by
equations (\ref{small_D_pert2})) that is formally analogous to that of
adiabatic perturbations.
This situation holds for whatever form of the contrast functions, as long
as these adimensional
quantities are very small (so that the initial functions
$\rhoim,\rhoir,{}^{(3)}R_i$ are almost
constant). Notice that, at first order expansion on the contrast functions,
we have for
(\ref{eqT})
\be \Psi \ \approx \
\left[\frac{\Psi}{\Gamma}\right]^{1/4} \ \approx \
 \ 1+\frac{1}{3}\,\delta\quad\quad \Rightarrow\quad \quad  T \ \approx \
\frac{T_i}{y}\,\left[\,1 +
\frac{1}{3}\,\delta\,\right],\label{eqT1}\ee
so that both, the Stefan-Boltzmann and Ideal Gas laws, yield the same
expression for $T$.
Therefore, as long as we assume the small contrasts approximation, the
Stefan-Boltzmann and Ideal
Gas laws yield the same result and we can use them indistinctly.

The resemblance of (\ref{small_D_pert2}) to expressions characteristic of
adiabatic perturbations
is not surprising: the generalized LTB models under consideration (under a
thermodynamic
approach) have zero heat flux and so can be associated with thermodynamic
processes that are
{\it indeed} adiabatic, though irreversible (because of the shear
viscosity). The connection
between the contrast functions and adiabatic perturbations on a
radiation-matter `tight coupling'
is an interesting feature that deserves proper examination and will be
studied in a separate
paper. We will use this analogy as
a theoretical tool for
understanding the type of quasi-homogeneous evolution and for defining
suitable initial
conditions for the study of the thermodynamic consistency of the models.
\section{Adiabatic and quasi-adiabatic initial conditions.}
As shown in \cite{mixt}, the quantity $\Delta_i^{(s)}=\Dri-(4/3)\Dmi$
becomes, under the
assumptions (\ref{small_D1}), (\ref{small_D2}) and (\ref{small_D3}), a sort
of average change of
photon entropy per barion at the initial hypersurface. Hence, it is
convenient to rephrase initial
conditions in terms of this quantity. After some algebraic manipulation on
the explicit general
forms given in Appendix A for the functions $A,B,C$ defined by
(\ref{eqABC}), we find that the
following restrictions on the initial contrast functions
\be \frac{4}{3}\Dmi \ = \ \Dri-\Delta_i^{(s)},\quad
\Dki \ = \ \frac{1}{2}\Dri,\label{initconds}\ee
leads to the following compact forms for the function $\Gamma$ defined in
(\ref{eqGamma1})
\ba \Gamma \ & = & \ 1 \ + \ 
\frac{3}{4}\left[1-\frac{\sqrt{q}}{y^2\sqrt{q_i}}\right]\,\Dri \ + \nonumber \\
             &   & \ \frac{3}{4y^2}\left[8\epsilon_1^2+4\epsilon_1y-y^2-
                   \frac{\sqrt{q}}{\sqrt{q_i}} \left(8\epsilon_1^2+
                   4\epsilon_1-1\right)\right]\,\Delta_i^{(s)}, \nonumber \\
             &   & \hbox{for the parabolic case} \quad (\epsilon_2=0), \nonumber \\
\Gamma  \    & = & \  1 \ + \
\frac{3}{4}\left[1-\frac{\sqrt{q}}{y^2\sqrt{q_i}}\right]\,\Dri \ + 
                     \frac{9}{4\epsilon_2\lambda_0 y^2}\nonumber \\
             &   & \left[\epsilon_1\left(1-
                   \frac{\sqrt{q}}{y^2\sqrt{q_i}}\right)-(\lambda_0\pm1) 
                   \left(y-\frac{\sqrt{q}}{\sqrt{q_i}}\right)\pm \frac{\lambda_0
                   (\eta-\eta_i)\sqrt{q}}{\sqrt{|\epsilon_2|}}\right]\,\Delta_i^{(s)}, \nonumber \\
             &   & \hbox{elliptic}\ (\epsilon_2>0,\,+\,\hbox{sign}) \ \hbox{and} \
                   \hbox{hyperbolic}\ (\epsilon_2< 0,\,-\,\,\hbox{sign})\
                   \hbox{cases},\label{eqGamma4}\ea
where $\epsilon_1=\omega/\mu,\,\epsilon_2=\kappa/\mu$ where defined in
(\ref{eqeps}) and
\ba q \ = -\epsilon_2\,y^2+2\,y+\epsilon_1,\quad q_i \ =
-\epsilon_2+2+\epsilon_1,\cr\cr \lambda_0 \ = \ 1\pm
\epsilon_1|\epsilon_2|,\quad
\hbox{elliptic}\, (+\,\,\hbox{sign}), \quad \hbox{hyperbolic}\,
(-\,\,\hbox{sign}),\cr\cr\cr
\eta \ = \
\arccos\left(\frac{1-\epsilon_2\,y}{\sqrt{\lambda_0}}\right),\quad \eta_i
\ = \
\arccos\left(\frac{1-\epsilon_2}{\sqrt{\lambda_0}}\right),\quad \hbox{elliptic
case},\cr\cr\cr \eta  = 
\hbox{arccosh}\left(\frac{1+|\epsilon_2|\,y}{\sqrt{\lambda_0}}\right),\
\eta_i  = 
\hbox{arccosh}\left(\frac{1+|\epsilon_2|}{\sqrt{\lambda_0}}\right),\
\hbox{hyperbolic case},\nonumber\ea
It is evident, by looking at these forms for $\Gamma$ above, that this
function takes a very
simple form
\be \Gamma \ = \
1 \ + \
\frac{3}{4}\left[1-\frac{\sqrt{q}}{y^2\sqrt{q_i}}\right]\,\Dri,\label{eqGamma_ad
}\ee
valid for all cases (parabolic, elliptic and hyperbolic) if we use initial
conditions
given by
\be  \Delta_i^{(s)} = 0 \quad \Rightarrow \quad \Dmi =
\frac{3}{4}\Dri,\quad \Dki =
\frac{1}{2}\Dri,\label{ad_initconds}\ee
Since all state variables, kinematic parameters and auxiliary functions are
constructed from
$\Gamma$, $y$ and initial value functions, the assumption
(\ref{ad_initconds}) leads to very
simplified forms for all expressions. Considering the initial contrast
functions
$\Dmi,\Dri,\Dki$ in (\ref{eqDeltas2}) as formally analogous to exact
perturbations on
$\rhoim,\rhoir,\Ri $, the factor $4/3$ relating $\Dmi$ and $\Dri$ in
(\ref{ad_initconds}) is
reminiscent of adiabatic perturbations on the initial value functions.
Hence, following
\cite{mixt} and \cite{susspav00}, we will denote (\ref{ad_initconds}) as ``adiabatic initial
conditions'', while the more general case (\ref{initconds}) with
$\Delta_i^{(s)}\ne 0$ will be refered to as ``quasi-adiabatic initial
conditions''. For
the remaining of this paper we will only consider initial conditions of
either these two
types, under the ``small contrast approximation'' given by (\ref{small_D1}),
(\ref{small_D2}) and (\ref{small_D3}).
\section{thermodynamic Consistency}
In order for the models presented in this paper to be physically meaningful
they must satisfy
energy conditions and must be compatible with causal Extended Irreversible
Thermodynamics.
Before to start this analysis we need first to provide an expression for the
phenomenological quantities $\eta,\,\alpha,\,\tau$ (coefficient of shear
viscosity, relaxation
time). Following the approach used in \cite{mixt} and \cite{susspav00}, we
consider for a
matter-radiation mixture interacting via radiative processes, the
``radiative gas'' model,
associated with the photon-electron interaction, which provides the
following forms for
$\eta$ and $\alpha$ \cite{wei} -\cite{JCL}
\be  \eta\ = \ {4\over{5}}\,p\,\tau \ = \ {4\over{15}}\,a\,T^4\tau,\quad
\alpha \ = \ -\frac{\tau}{2\eta} \ = \
-\frac{15}{8\,a\,T^4},\label{phen_coefs}\ee
Inserting (\ref{phen_coefs}) into (\ref{eqs1}) and (\ref{eqtr1}), the latter
equations become
\be\dot s \ = \ \frac{5\,k}{8\,\tau}\,\frac{\Pi_{ab}\Pi^{ab}}{p^2}
\ = \ \frac{15\,k}{4\,\tau}\,\frac{P^2}{p^2},\label{eqsdot1}\ee
\be \dot P+\left(\frac{4}{3}\Theta+\frac{1}{\tau}\right)
P+\frac{8}{5}\,p\,\sigma\left[1+\nu_0\,\left(\frac{P}{p}
\right)^2\right] \ = \ 0, \label{eqtr2}\ee
where in (\ref{eqtr2}) $\nu_0=5/4,\,25/32$, respectively, for the ideal gas
and Stefan-Boltzmann
laws (we examine the truncated equation (\ref{eqtrmc1}) further ahead). The
transport equation
(\ref{eqtr2}) is an evolution equation for $P$ imposed by thermodynamic
theories external (even if coupled) to General Relativity.  On the other
hand, there are evolution
equations for $P$ that follow from the field equations, for example, by
evaluating $\dot P$ from
(\ref{eqrhopP}) and eliminating $\dot Y/Y$ and $\dot Y'/Y'$ with the help
of (\ref{eqTheta1}) and
(\ref{eqsigma1}), leading to
\be \dot P \ + \ \left(5\sigma + \frac{4}{3}\Theta\right) P \ - \
2\,p\,\sigma \
=
\ 0, \label{dotP_FE}\ee
an exact equation that must be satisfied by the generalized LTB models.
Obviously, an equation
like (\ref{dotP_FE}) does not exactly coincide with (\ref{eqtr2}), and so
compatibility between
these models and Extended Irreversible Thermodynamics requires finding an
appropriate expression for $\tau$ that should make (\ref{dotP_FE})
consistent with the transport
equation. Comparing (\ref{eqtr2}) (with $\nu_0=5/4$, ideal gas) with the
evolution equation
(\ref{dotP_FE}), we can see that both equations coincide if we identify
\be \tau  =  \frac{-p\,P}{\sigma\,\,\left\{
2\,P^2+\frac{18}{5}\,p^2-5\,p\,P\right\}}
 =  -{\frac{1}{4\sigma}}\;
\frac{\left[4+3\Delta_i^{^{(r)}}-4\Gamma\right]\left[4+
3\Delta_i^{^{(r)}}-\Gamma\right]}{{\textstyle{{4} \over
{5}}}\left[4+3\Delta_i^{(r)}+{\textstyle{{13} \over
{32}}}\Gamma\right]^2+{\textstyle{{171}
\over {256}}}\Gamma^2 },\label{eqtau}
\ee
This expression is justified as long as it behaves as a relaxation parameter
for the interactive matter-radiation mixture in the theoretical framework of
EIT as we will explain bellow.

It is also useful to compute the collision times for Thomson and Compton
scattering
(the dominant radiative processes in the radiative era)

\be t_{_{\gamma}}=\frac{1}{c\sigma_Tn_e},\quad t_{c}=\frac{m_e
c^2}{k_{_{B}}T}\,t_{_{\gamma}}\, ,\label{eqtcoll}
\ee
where $\sigma_T$ is the Thomson scattering cross section, $m_e$ is the
electron mass and $n_e$ is
the number density of free electrons, a quantity obtained from Saha's
equation, leading to
\be
t_\gamma ={1 \over {2c\,\sigma _{_T}n^{(m)}}}\left[ {1+\left(
{1+{{4h^3n^{(m)}\exp \left( {B_0/k_{_B}T} \right)} \over {\left( {2\pi
\,m_ek_BT} \right)^{3/2}}}} \right)^{1/2}} \right], \label{eqtgama}
\ee
\noindent
where $B_0$ and $h$ are respectively the hydrogen atom binding energy
and Planck's constant. For details of the derivation of (54) see
\cite{mixt}.

The restrictions for physical acceptability and thermodynamic
consistency were discussed in \cite{susspav00} and \cite{mixt}. We
provide a summary in the following list:

\bi

\item
(i) Positive definiteness and monotonicity of $\rhor$ and $\rhom$, and $|P|
\ll p$.

\item
(ii) Regularity condition
\be \Gamma \ > \ 0,  \ee
which prevents the occurrence of a shell crossing singularity \cite{singl}.

\item
(iii) Positive definiteness of $\tau$ and $\dot s$ (consistent with
positive entropy production). The sign of $\dot s$ (see (\ref{eqsdot1}))
depends only
on the sign of $\tau$.

\item
(iv) Concavity and stability of $s$ by requiring
\be \dot \tau>0,\quad {\ddot s\over\dot
s}= {{2\sigma\Gamma}\over{3\Psi\Phi}}\,{{\left\langle
\rhoir\right\rangle}\over
{\rhoir}}\left[1+{{\left\langle \rhoir\right\rangle}\over
{3\rho_i^{(r)}}}\right]-{\dot\tau\over{\tau}} < 0.
\ee

\item
(v) Appropriate behavior of the relaxation time $\tau$.

During the radiative era the Thompson time $t_{\gamma}$ must be
smaller than the expansion (or Hubble) time, approximately defined by
$t_{H}=3/\Theta$. The decoupling hypersurface is then defined by
$t_{\gamma}=t_{H}$. From (iii) and (iv) above \cite{susstrig}, \cite{mixt},
$\tau$  must be a
positive and monotonously increasing function (if the fluid expands), then
it must also be (during
the interactive period) qualitatively similar but larger than the
microscopic timescales of the
various photon-electron interactions occurring in the radiative era. It is
usually assumed
that $\tau$ is of the order of magnitude of a collision time, and as matter and
radiation decouple all these timescales must overtake the Hubble expansion time
$t_{_H}=3/\Theta $. A physically reasonable $\tau$ should be
comparable in magnitude to $t_{\gamma}$ near decoupling and must have
an analogous qualitative behavior to $t_{\gamma}$. Therefore $\tau$
must be smaller than $t_{H}$ during the radiative era and then must
overtake it at the decoupling hypersurface.

\ei

In the following list we test the previous restrictions (i) to
(v) under the assumption of the small contrasts approximation (\ref{small_D1}),
(\ref{small_D2}) and (\ref{small_D3}), together with either adiabatic
(\ref{ad_initconds})
or quasi-adiabatic (\ref{initconds}) initial conditions:

\bi
\item
(i) For a wide range of initial conditions all these quantities have a
physically meaningful
behavior as indicated in Figure.~\ref{FrhoP}.

\item
(ii) Figure.~\ref{FGamma} shows that $\Gamma$  is
positive and almost equal to unity for most  of the evolution range of $y$.
For the small range
defined by $10^{-5} \le {|\Delta_{i}^{(S)}|} \le 10^{-3}$ and $\log_{10}(y)
\le 1$ the
plot of $\Gamma$ is distinct for lumps and voids.

\begin{figure}
\caption{Plot of $\log_{10}(\tau/t_{_{H}})$ in terms of
$\log_{10}(y)$, (a) for adiabatic conditions $|\Delta_{i}^{(s)}|=0$,
$\tau/t_{_{H}}$ is always smaller then 1, so $\tau$ does not overtake
$t_{_{H}}$, (b) For $\tau$ defined from the truncated equation (49c)
$\tau/t_{_{H}}$ is always smaller then 1 as well.} \label{FtautHLast}
\end{figure}
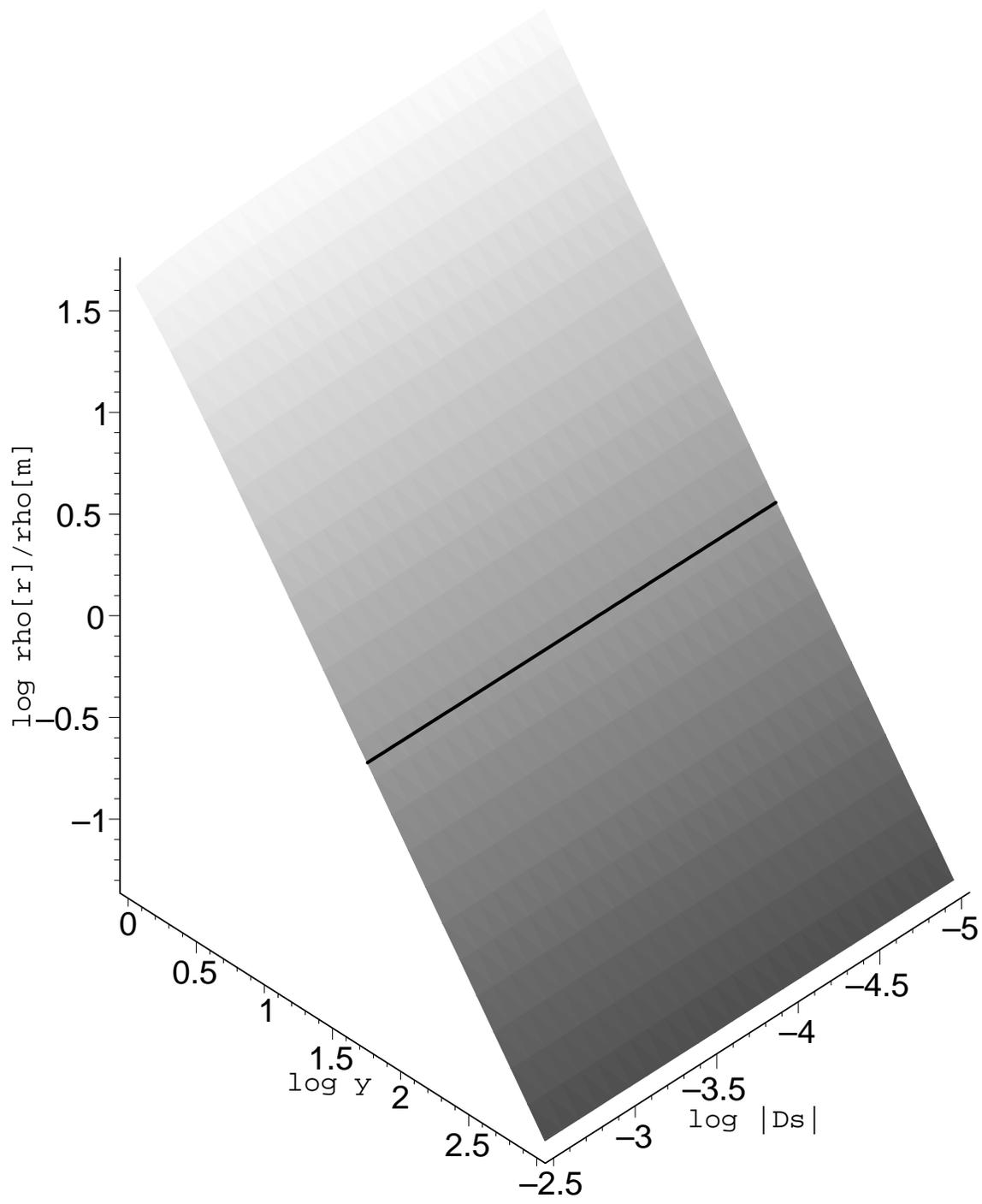
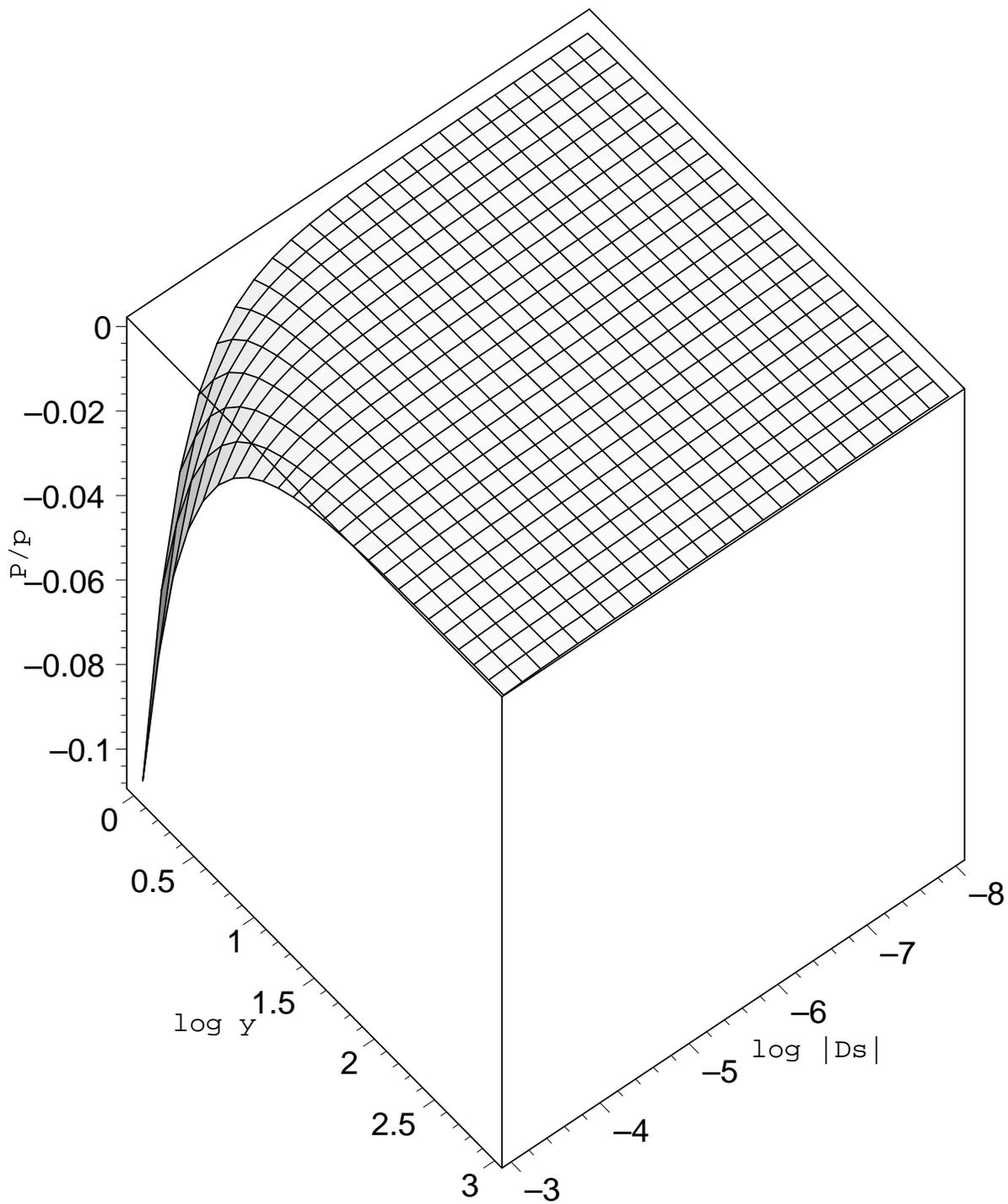
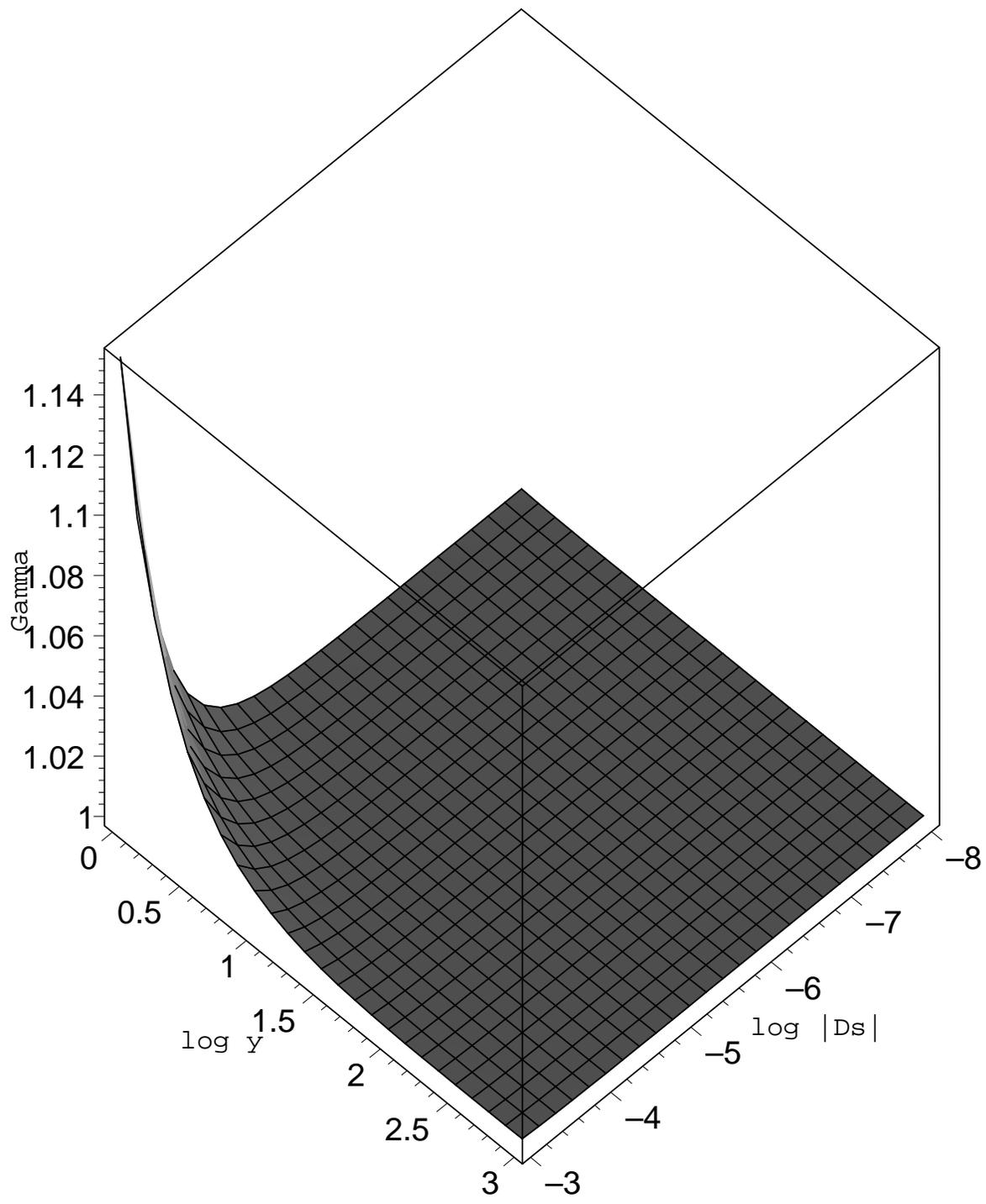
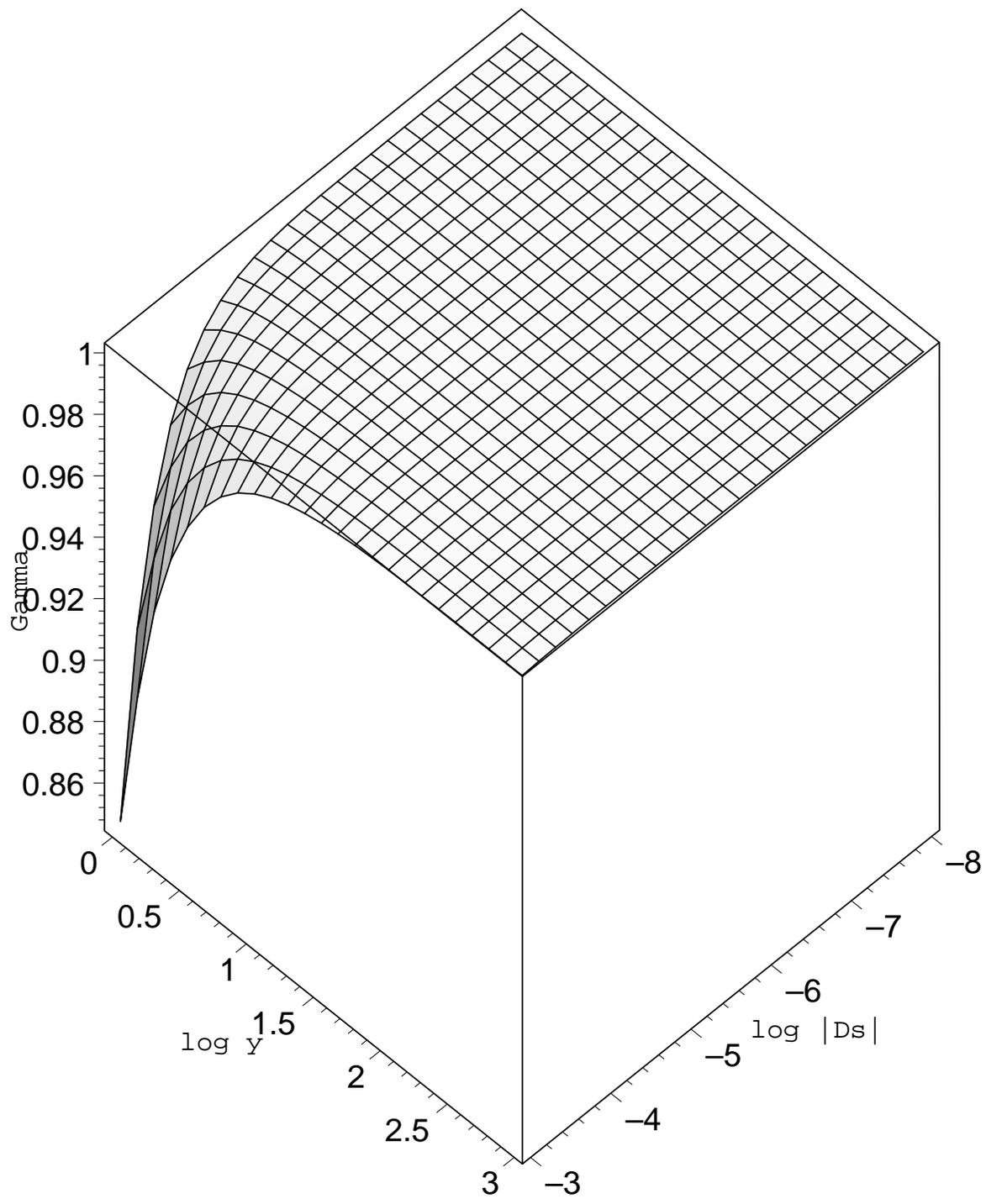
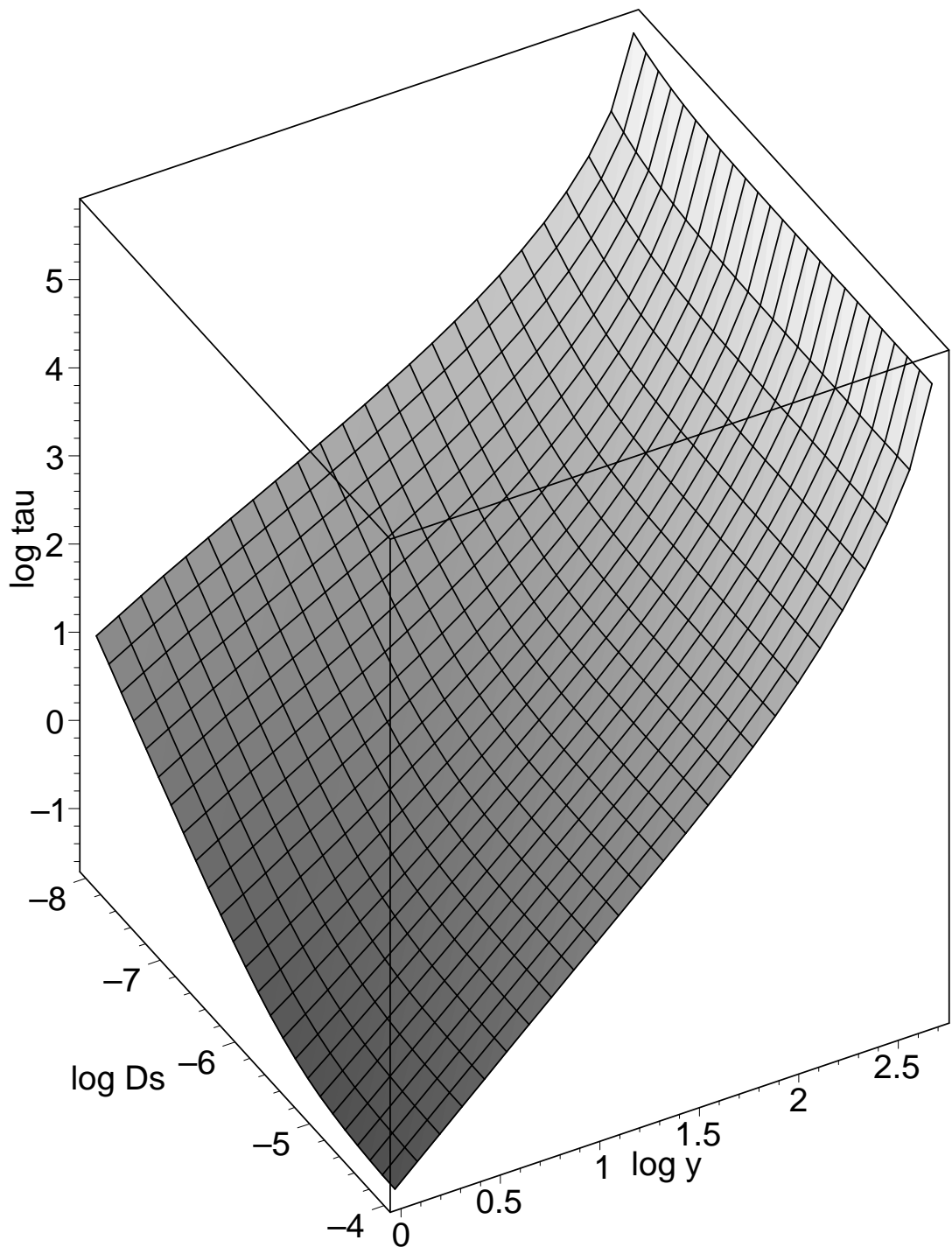
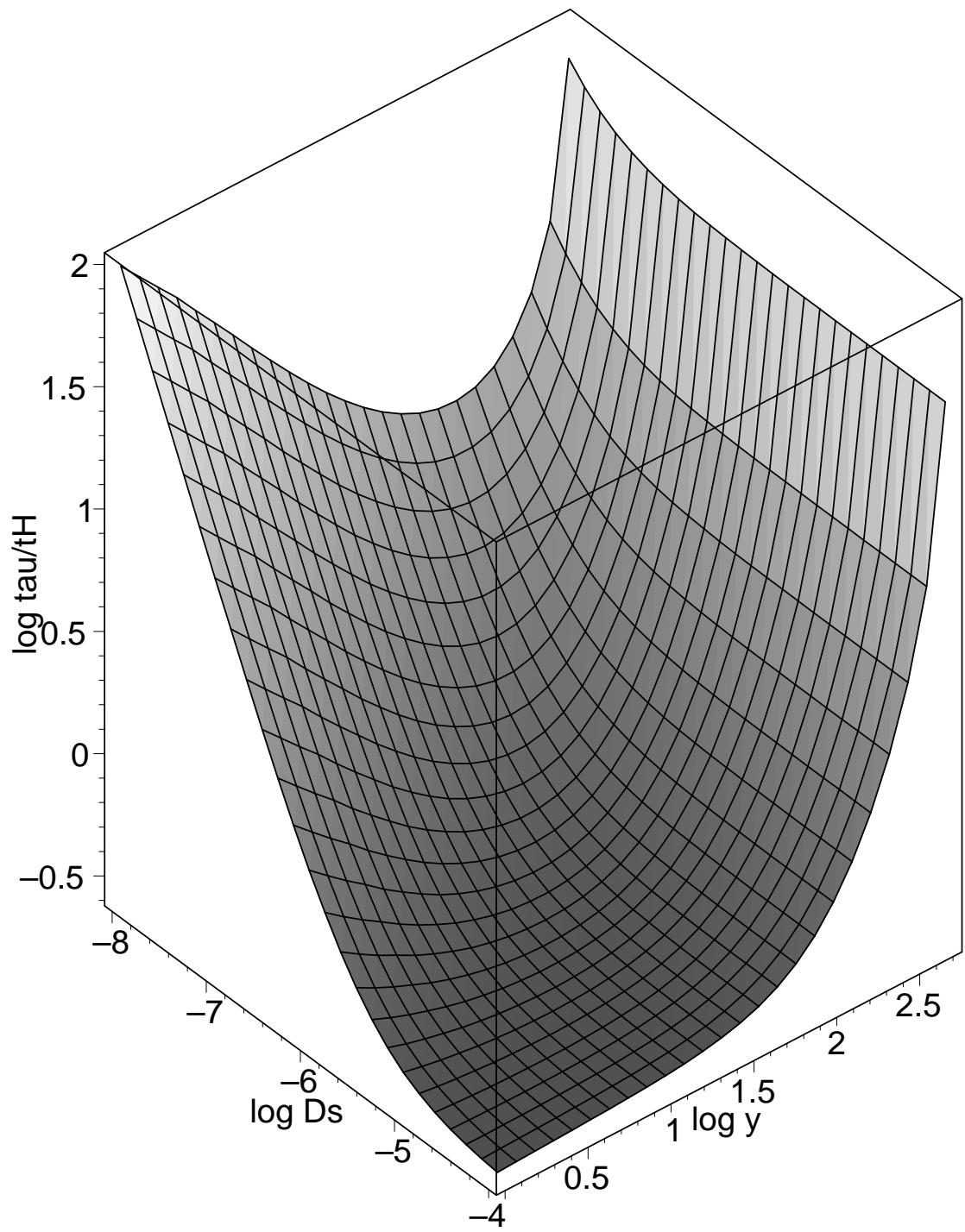
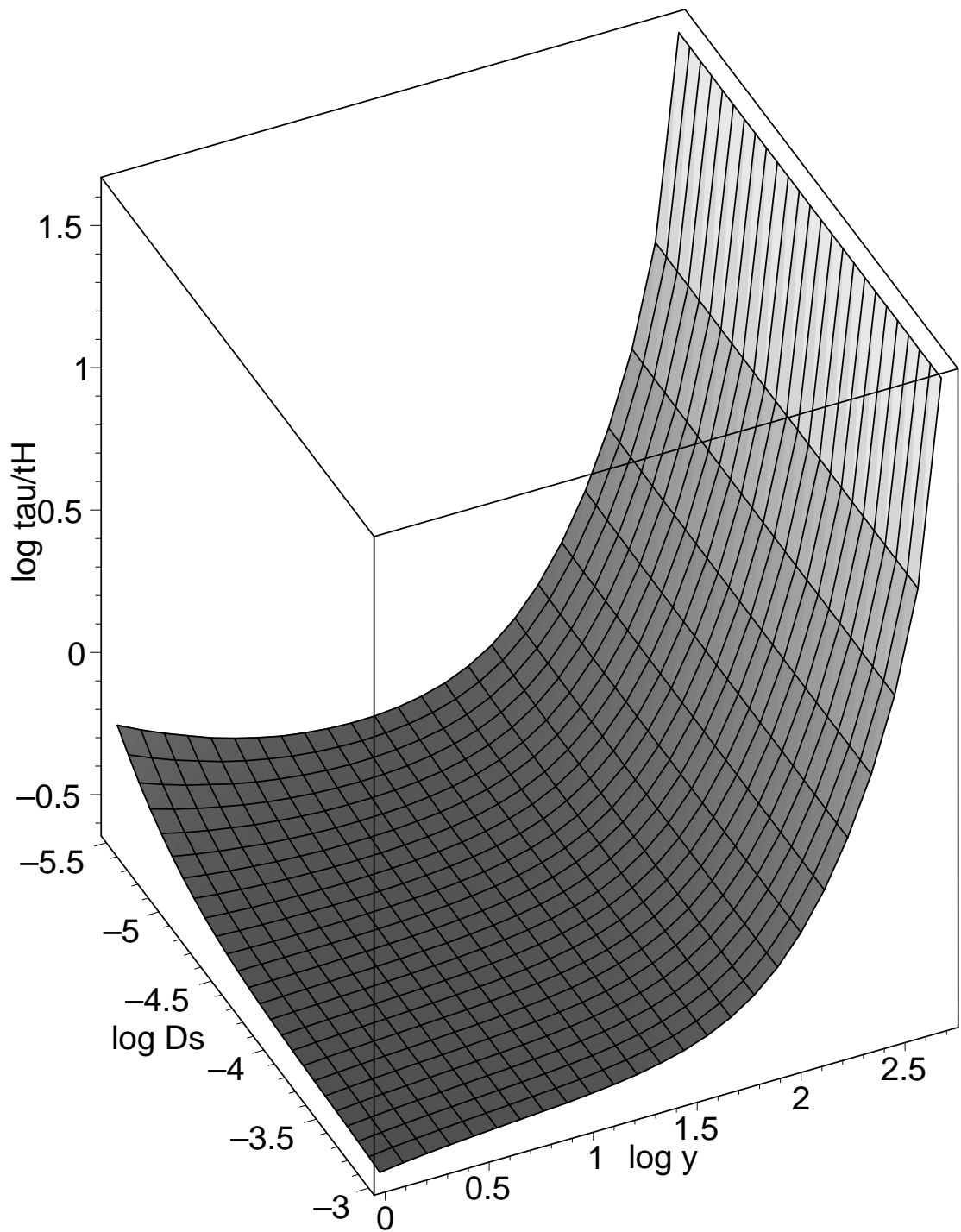
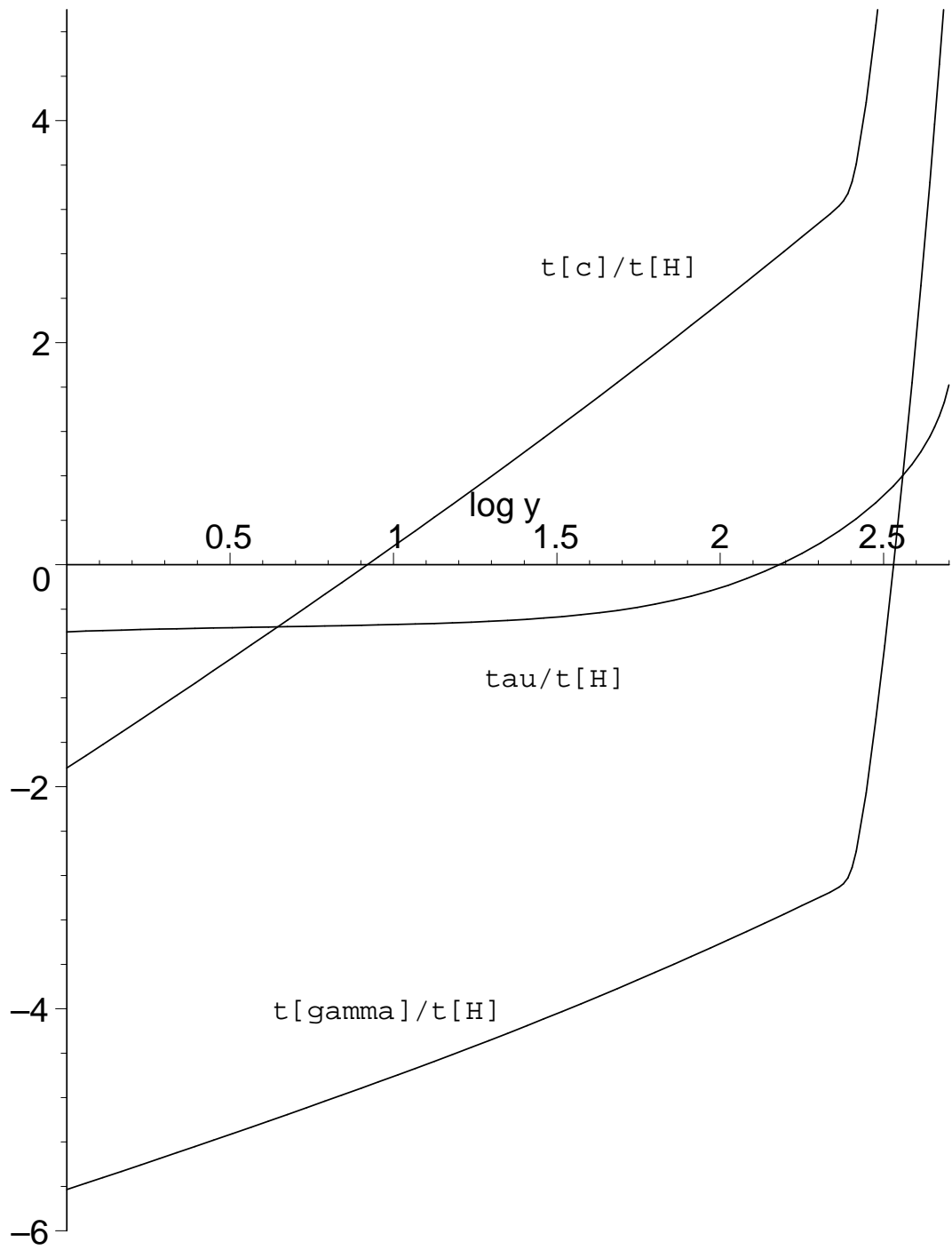
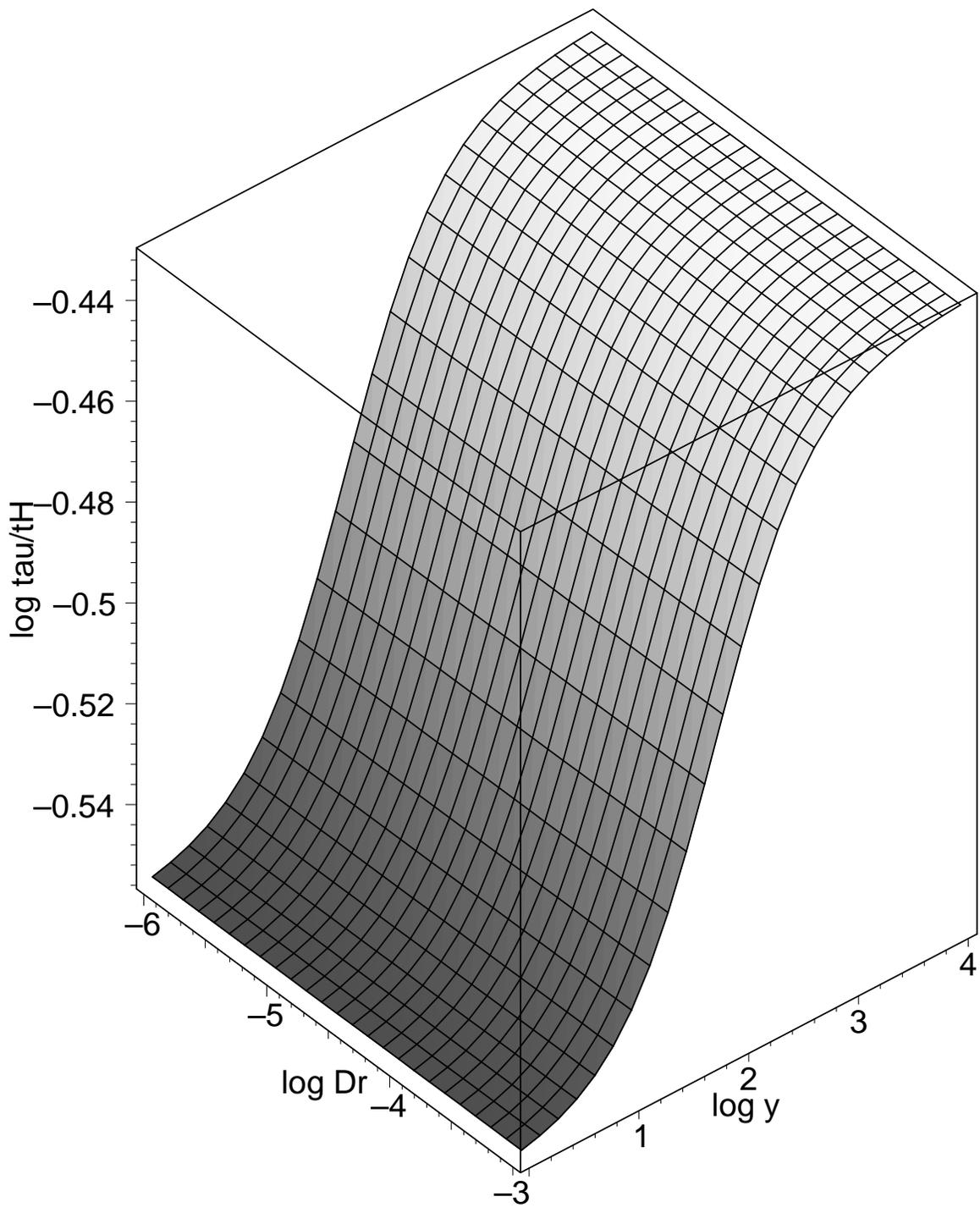
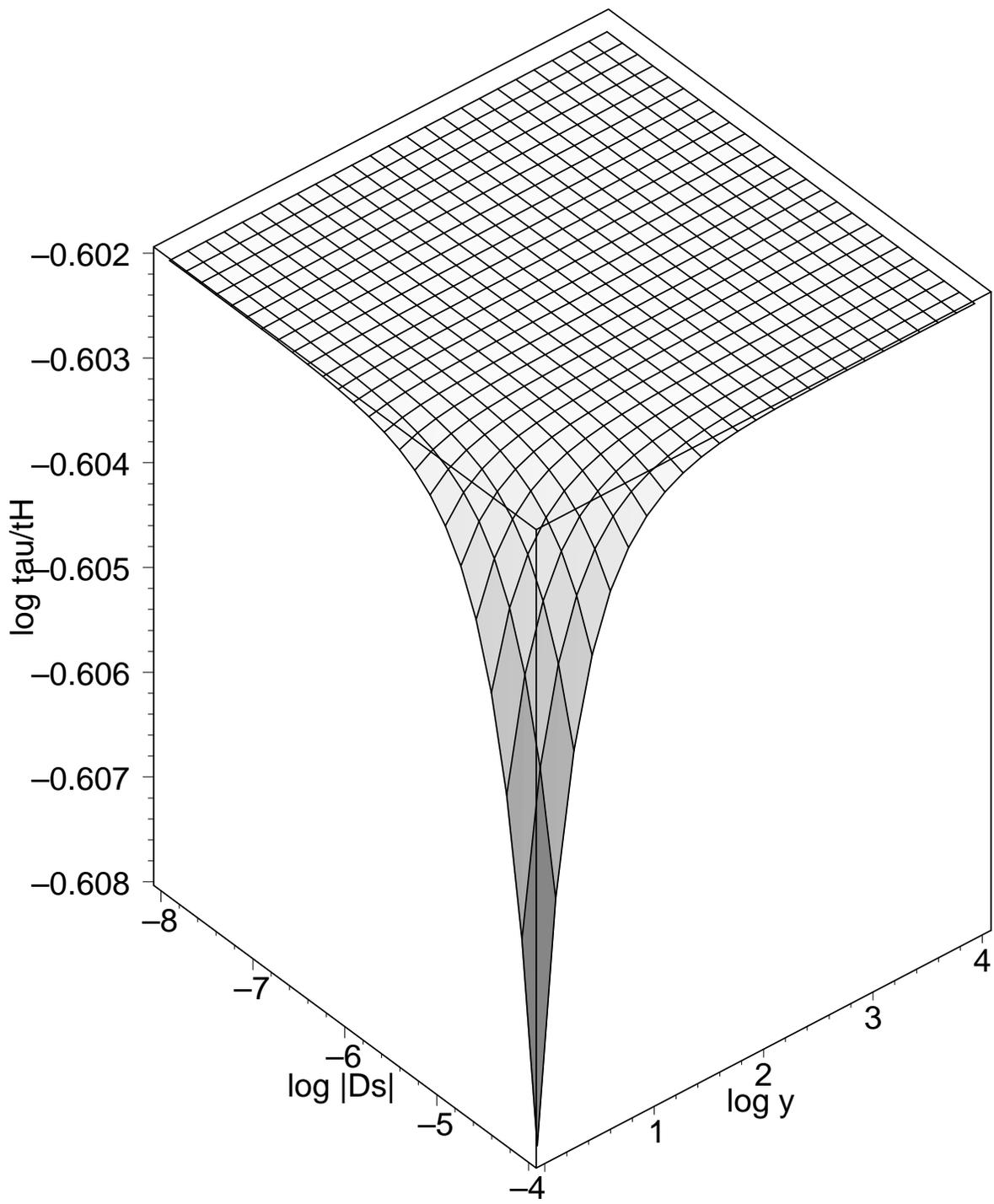

\item
(iii) and (iv). Figure.~\ref{FFtau} shows that
$\tau>0,\,\dot s>0$ are satisfied for quasi-adiabatic initial conditions.
From Figure.~\ref{FFtau} it can
also be seen that  $\tau$ is a monotonically increasing function and so
$\ddot s<0 $. The same
results hold for adiabatic conditions (see Figure.~\ref{FtautHLast}a).

\item
(v) Under adiabatic initial conditions ($|\Delta_{i}^{(s)}|=0$), $\tau$
does not have the required
behavior described above (never overtakes $t_{_{H}}$). This is shown in
Figure.~\ref{FtautHLast}a.  We find that the desired behavior of $\tau$ is
encountered for
quasi-adiabatic initial conditions. As shown in Figure.~\ref{FtautH}, this
happens in the elliptic case for the values
$10^{-5.5}\le|\Delta^{(s)}|\le10^{(-3)}$, compared to
$|\Delta^{(S)}|\le10^{(-8)}$ in the hyperbolic (not shown) and parabolic
case \cite{susspav00}.
Finally, we plot the ratios $\tau/t_{_{H}}$, $t_c/t_{_{H}}$ and
$t_{\gamma}/t_{_{H}}$ in
Figure.~\ref{Ftautgammatc} showing that these times have a physically
reasonable and consistent
behavior.
\ei

Furthermore, we examine the truncated transport equation for the generalized LTB models that follows by
inserting
(\ref{phen_coefs}) into (\ref{eqtrmc1})
\be \dot P+
\frac{1}{\tau}\,P+\frac{8}{5}\,p\,\sigma=0,\label{eqtrmc2}\ee
A comparison between (\ref{eqtr2}) and (\ref{eqtrmc2}) shows that the full
equation is
equivalent to the truncated one if $P/p\ll 1$ and $1/\tau\ll 4\Theta/3 $.
The first condition
simply requires small deviations from equilibrium (as shown by
(\ref{eqsdot1})) and is
compatible with energy conditions \cite{susstrig}.
However, the condition $4\tau\gg 3/\Theta$ is more problematic and is only
reasonable after
matter and radiation have decoupled, hence, under the assumptions
underlying the thermodynamic
study of the models, the truncated equation cannot describe the interacting
period nor the
decoupling process. This shortcoming of (\ref{eqtrmc1}) has been discussed
previously for the
parabolic case \cite{susspav00}. As shown by Figure.~\ref{FtautHLast}b,
this situation holds also
for the elliptic and hyperbolic cases.

In order to complete the analysis we compute the Jeans mass associated to
the initial
conditions of the case $\Delta_i^{(s)}\ne 0$. This mass is given
by\cite{sw},\cite{bor},\cite{pad}

\be 
M_{_J}=\frac{4\pi}{3}m\,n^{(m)}\left[\frac{c^4 \pi
C_s^2}{G\,(\rho+p)}\right]^{3/2}=
 \frac{4\pi}{3}\frac{c^4\chi_i\Gamma^{1/2}}{\sqrt{\rhoir}}
\left[\frac{\pi y^2\Psi}{3G\left(\Psi+\textstyle{{3} \over
{4}}\chi_iy\right)^2}\right]^{3/2}, \label{jeansmass}
\ee
\noindent where $\rho,\,p,\,n^{(m)}$ are given by (\ref{eqst2}),
(\ref{eqstig}) and (\ref{small_D2}),
$\chi_i=\rho_i^{m}/\rho_i^{r}$, $\Psi$ and $\Gamma$ follow from (\ref{eqGamma1}) and
$C_s$ is the speed of sound, which for the equation of state (\ref{eqst2}), has the form

\be 
C_s^2=\frac{c^2}{3}\left[1+\frac{3\rho^{(m)}}{4\rho^{(r)}}\right]^{-1},\quad
\rhom=mc^2n^{(m)}, \quad \rhor=3n^{(r)}k_{_B}T.
\ee

\noindent Evaluating (\ref{jeansmass}) for $y=y_{D}\approx 10^{2.4}$,
$\epsilon\approx 1/\chi_i\approx 10^3$ and $\rhoir\approx
a_{_B}T_i^4\approx 7.5\times
10^9\,\hbox{ergs}/\hbox{cm}^3$, yields $M_{_J}\approx
10^{49}\hbox{gm}$, or approximately
$10^{16} M\odot$. This value coincides with the Jeans mass obtained
for perturbative models dominated by baryons in the radiative era as
decoupling is approached.
Finally, it is worthwhile remarking that we are dealing with an epoch where a
cosmological constant or a quintessence component should be clearly
sub-dominant \cite{cosmoquint} and ignored.
\section{Conclusion}
We have derived and discussed important generic properties of a class of exact
solutions of Einstein's equations that generalize the famous LTB solutions
with a dust source to an imperfect fluid with anisotropic pressure $\Pi_{ab}$.
The integration of the field equations does not involve $\Pi_{ab}$,
though once this
integration is done this pressure becomes also determined (up to initial
conditions). The issue regarding the compatibility of $\Pi_{ab}$ and
its evolution
law with the physical assumptions underlining the models have been
addressed. These models provide a physically plausible hydrodynamical
description of cosmological matter-radiation mixture in the radiative
era, between nucleosynthesis and decoupling. By assuming small initial
density contrasts (consistent with small deviations from equilibrium), we have
shown that the state variables of the models are qualitatively and formally
analogous to that of adiabatic perturbations on a FLRW background.  We have
also
found two classes of initial conditions, based on well defined initial contrast
functions, that are formally equivalent to the definition of adiabatic and
quasi-adiabatic
perturbations on initial value functions $\rhoim,\,\rhoir,\,\Ri$. In
particular, we showed
that for quasi-adiabatic initial conditions the models are
thermodynamically consistent
and physically acceptable. However, this consistency does not hold for
the truncated
transport equation (as shown also in \cite{susspav00}), only for the full
transport equation of
Extended Irreversible Thermodynamics the relaxation time of shear viscosity
has the appropriate
physical behavior.
\acknowledgements
This work was partially supported by the National University of Mexico
(UNAM) under grant DGAPA-IN-122498 (to RS), Ontario Graduate Scholarship
for Science and Technology and Ontario Graduate Scholarship to (MI). 
Portions of this work were made possible by use of GRTensorII\cite{grtensor}.
\begin{appendix}
\section{Integration of the Friedmann equation (22).}
The quadrature (\ref{ydot}) depends on the sign of $\kappa=K/Y_i^2$. The
integral yields
either an implicit function $ct= Z(y,\mu,\omega,\kappa)$ where $Z$ is
defined by (\ref{eqZ})
(the ``canonical solution''), or a parametric solution having the form
$y=y(\eta,r),\,t=t(\eta,r)$. The canonical solution of the equation
equivalent to
(\ref{ydot}) found in the literature for the LTB dust case
\cite{ksmh},\cite{kras}, is usually
given in the form $c[t-t_0(r)]= Z(y,\mu,\kappa)$, where the function
$t_0(r)$ comes as an
integration constant. Instead, we will provide the canonical solutions in
the form
$c(t-t_i)=Z-Z_i$, so that we can identify $t_0(r)=t_i+Z_i$. The function
$t_0(r)$ can also be
identified as the ``big bang time''.  Only for $\kappa=0$ the canonical
solution can be inverted
as $y=y(t,r)$ by solving a cubic equation, however the resulting expression
is too cumbersome and
will not be given.

All variables introduced in previous sections will be made fully
determined as functions of $y$ and of $r$, with the dependence on $r$ given
through
the initial value functions and their initial contrast functions.  As usual
in the study
of dust LTB solutions, we will classify the integrals (\ref{eqZ}) in three
cases:
``parabolic'' ($\kappa=0$), ``elliptic'' ($\kappa>0$) and ``hyperbolic''
($\kappa<0$). We shall
examine each case separately below.
\subsection{Parabolic case.}
If $\kappa=0$, so that $\Ri=\Dki=0$, then (\ref{eqZ}) leads to the
canonical form
\be c(t-t_i)=Z-Z_i= \frac{4}{3\sqrt{\mu}}\left[\sqrt{2y+\epsilon_1}\,(
y-\epsilon_1)-\sqrt{2
+\epsilon_1}\,(1 -\epsilon_1)\right]\ee
where $\epsilon_1$ was defined in (\ref{eqeps}). The solution in terms of
the parameter $\eta$ is
\be y \ = \ \frac{1}{2}\left[\frac{\eta}{\mu}-\epsilon_1\right],\quad
c(t-t_i) \ = \
\frac{\sqrt{\eta}}{6\mu}\left[\frac{\eta}{\mu}-\epsilon_1\right] \ - \ Z_i\ee
The functions $A,\,B$ in (\ref{eqABC}) have the form
\ba A \ = \
\frac{1}{3y^2}\left[\frac{\sqrt{2y+\epsilon_1}}{\sqrt{2+\epsilon_1}}
(1-4\epsilon_1-8\epsilon_1^2)-(y^2-4\epsilon_1
y-8\epsilon_1^2)\right],\cr\cr\cr
B \ = \ \frac{2\epsilon_1}{y^2}\left[(1+\epsilon_1)\frac{\sqrt{2y+\epsilon_1}}
{\sqrt{2+\epsilon_1}}-(y+\epsilon_1)\right]\ea
Notice that the limit $\rhoir\to 0$ (or $\epsilon_1\to 0$) leads to the known
parabolic solutions of the dust LTB solutions, though, the limit $\rhoim\to
0$ is
singular.
\subsection{Elliptic case}
For $\kappa>0$ the canonical integral (\ref{eqZ}) is

$$c(t-t_i)=Z-Z_i=\left|-\frac{\sqrt{Q}}{\kappa}+\frac{\mu}{\kappa^{3/2}}\arccos
\left (\frac{Q_{,y}} {\sqrt{\lambda}}\right)\right|_{y=1}^{y}\eqno(A5)
$$
$$Q=-\kappa y^2+2\mu y+\omega,\quad Q_{,y}=-2\kappa y+2\mu,\quad
\lambda\equiv
4(\mu^2+\kappa\omega)
$$
while the parametric solution is given by
$$y=\frac{\mu}{\kappa}\left[1-\frac{\sqrt{\lambda}}{2\mu}\cos\,\eta\right]
\eqno(A6a)$$
$$c(t-t_i)=\frac{\mu}{\kappa^{3/2}}
\left[\eta-\frac{\sqrt{\lambda}}{2\mu}\sin\,\eta\right]-Z_i\eqno(A6b)
$$
The functions $A,\,B,\,C$ in (\ref{eqABC}) become
$$A=\frac{\mu}{\kappa
y^2}\left[\frac{(\eta_i-\eta)\sqrt{Q}}{\sqrt{\kappa}}+\frac{4\mu^2+
\lambda+4
\mu\omega}{\lambda}\,\frac{\sqrt{Q}}{\sqrt{Q_i}}-\frac{(4\mu^2+\lambda)y+4
\mu\omega}{\lambda}\right]\eqno(A7a)
$$
$$B=\frac{2\omega}{\lambda
y^2}\left[\left(\omega+\mu\right)\frac{\sqrt{Q}}{\sqrt{Q_i}}-\left(\omega
+\mu y\right)\right]\eqno(A7b)
$$
$$ C=-\frac{3\mu\sqrt{Q}(U_i-U)}{2\kappa^{3/2}y^2}+\frac{\left[(\lambda-4\mu\omega)\kappa-
(2\mu+\omega)(4\mu^2+2\lambda)\right]\sqrt{Q}}{2\kappa\lambda
y^2\,\sqrt{Q_i}}-$$
$$ \frac{\left[(\lambda-4\mu\omega)\kappa
y-(2\mu y+\omega)(4\mu^2+2\lambda)\right]\sqrt{Q_i}}{2\kappa\lambda
y^2\,\sqrt{Q_i}}\eqno(A7c)
$$
$$\eta = \arccos\left(-\frac{Q_{,y}}{\sqrt{\lambda}}\right) $$
As with the parabolic case, we obtain the dust LTB elliptic solution in the
limit $\omega\to 0$. It is also interesting to compare the evolution of $y$ as
function of $\eta$ with the dust case. This evolution is also
time-symmetric, but it
begins at the value $\eta_{BB}=\arccos(2\mu/\sqrt{\lambda})>0$, instead of
$\eta_{BB}=0$. The maximal value of $y$ as the fluid bounces is
$y_{max}=(\mu/\kappa)(1+\sqrt{\lambda}/2\mu)$, larger than the value for
the dust
case $y_{max}=(\mu/\kappa)$.
\subsection{Hyperbolic case}
For $\kappa <0$ the canonical integral (\ref{eqZ}) is

$$c(t-t_i)=Z-Z_i=\left|\frac{\sqrt{Q}}{|\kappa|}-\frac{\mu}{|\kappa|^{3/2}}
\hbox{arccosh}\left(\frac{Q_{,y}}
{\sqrt{\lambda}}\right)\right|_{y=1}^{y},\quad \lambda>0\eqno(A8a)
$$
$$c(t-t_i)=Z-Z_i=\left|\frac{\sqrt{Q}}{|\kappa|}-\frac{\mu}{|\kappa|^{3/2}}
\ln\left(\frac{Q_{,y}}
{2\sqrt{|\kappa|}}+\sqrt{Q}\right)\right|_{y=1}^{y},\quad \lambda\leq
0\eqno(A8b)
$$
$$Q=|\kappa|y^2+2\mu y+\omega,\quad Q_{,y}=2|\kappa|y+2\mu,\quad
\lambda\equiv 4\mu^2-4|\kappa|\omega
$$
while the parametric solution is given by
$$y=\frac{\mu}{|\kappa|}\left[\frac{\sqrt{\lambda}}{2\mu}\cosh\,\eta-1\right]
\eqno(A9a)$$
$$c(t-t_i)=\frac{\mu}{|\kappa|^{3/2}}
\left[\frac{\sqrt{\lambda}}{2\mu}\sinh\,\eta-\eta\right]-Z_i\eqno(A9b)
$$
The functions $A,\,B,\,C$ in (14) become
$$A=\frac{\mu}{\kappa
y^2}\left[\frac{(\eta_i-\eta)\sqrt{Q}}{\sqrt{|\kappa|}}+\frac{4\mu^2+\lambda+4
\mu\omega}{\lambda}\,\frac{\sqrt{Q}}{\sqrt{Q_i}}-\frac{(4\mu^2+\lambda)y+4
\mu\omega}{\lambda}\right]\eqno(A10a)
$$
$$B=\frac{2\omega}{\lambda
y^2}\left[\left(\omega+\mu\right)\frac{\sqrt{Q}}{\sqrt{Q_i}}-\left(\omega
+\mu y\right)\right]\eqno(A10b)
$$
$$C=-\frac{3\mu\sqrt{Q}(\eta_i-\eta)}{2|\kappa|^{3/2}y^2}+\frac{\left[(\lambda-4
\mu\omega)\kappa-
(2\mu+\omega)(4\mu^2+2\lambda)\right]\sqrt{Q}}{2\kappa \lambda
y^2\,\sqrt{Q_i}}- $$
$$ \frac{\left[(\lambda-4\mu\omega)\kappa y
-
(2\mu y+\omega)(4\mu^2+2\lambda)\right]\sqrt{Q_i}}{2\kappa \lambda
y^2\,\sqrt{Q_i}}\eqno(A10c)
$$
$$\eta \ = \ \hbox{arccosh}\left(-\frac{Q_{,y}}{\sqrt{\lambda}}\right) $$
As with the parabolic case, we obtain the dust LTB elliptic solution in the
limit $\omega\to 0$.
\section{Particular cases.}
Expressing initial value functions in terms of the initial averages and
contrast functions defined in sections III and IV greatly facilitates the
description
of the correspondence to the following particular cases:
\begin{itemize}
\item {\bf Dust limit.}
If $\rhoir=\<\rhoir\>=0$, then $\omega=0$ and $Q=(-\kappa y+2\mu)y$,
hence we have: $\rho=\rhom=\rhoim/(y^3\Gamma)$ and $p=P=0$, and so the
solutions reduce to the ``usual'' LTB solutions with a dust source. The initial
density and curvature contrasts $\Dmi,\,\Dki$ have the same
interpretation as in the general case. Notice that $B=0$ (from (14a)) and
that the
auxiliary functions $\Psi,\,\Phi$ are irrelevant in this sub-case.

\item {\bf FLRW limit.}
Homogeneity along the initial hypersurface follows by demanding
$\Dmi=\Dri=\Dki=0$, leading to $\rhoim=\<\rhoim\>$, $\rhoir=\<\rhoir\>$,
$\Ri=\<\Ri\>$, so that
$\rhoim,\,\rhoir,\,\Ri$ are constants. Equation (\ref{ydot}) implies then that
$\mu,\,\omega,\,\kappa$ are also constants,  so that $M,\,W\propto Y_i^3 $
and $K\propto Y_i^2$.
Therefore $y$ obtained by integrating (\ref{ydot}) must be a function of
$t$ only and
$Y=R(t)Y_i$. From (\ref{eqGamma1}) to (\ref{eqPhi1}) we have:
$\Gamma=\Psi=1$ and $\Phi=0$
leading to $\rho=\rho(t),\, p=p(t),\,P=0$ and $\sigma=0,\,\Theta=3\dot
y/y=3\dot R/R$. Thus the
particular case in which initial density and curvature contrasts vanish is
the FLRW
limit of the solutions, a FLRW cosmology where $\rho,\,p$ satisfy the
``dust plus
radiation'' relation (\ref{eqrhop}).

\item {\bf Vaidya limit.}
If $\rhoim>0,\, \rhoir>0$ for $0\leq r< r_b$ but vanish for $r\geq r_b$,
then $\<\rhoim\>$ and $\<\rhoir\>$, as well as $M$
and $W$ are constants for $r\geq r_b$ and we have (along this range) a
sub-case of the
Vaidya metric characterized by the mass function $m=M(r_b)+W(r_b)Y_i/Y$.
This particular Vaidya solution is the ``exterior'' field for the models
derived in
this paper, generalizing the Schwarzschild exterior of LTB dust solutions.
Notice
that as the fluid expands the Vaidya mass function, $m$, tends to a constant
Schwarzschild mass given by $M(r_b)$.  More details are provided in
\cite{susstrig}.

\end{itemize}

The solutions derived in this paper can be smoothly matched along a comoving
hypersurface marked by $r=r_b$ with their FLRW and Vaydia sub-cases. The
matching
conditions are discussed in detail in \cite{susstrig}. Just as LTB dust
solutions
can be generalized to the Szekeres dust solutions without isometries, LTB
solutions with an imperfect fluid source can be generalized to the
Szafron-Szekeres
metrics with an imperfect fluid source\cite{susstrig}. Even without spherical
symmetry, these solutions can still be smoothly matched to the FLRW and Vaydia
sub-cases along a spherical comoving boundary \cite{susstrig}.
\end{appendix}
\begin{figure}
\caption{(a) Logarithmic plot of the densities ratio. These
densities are positive definite and decreasing
functions of $\log_{10}(y)$; The  plot also shows that for the early
stage the radiation density dominates while for later stages the
matter density dominates. (b) The ratio 
$P/p$ is very small complying with energy condition. $P$ is
negative/positive for lumps/voids respectively. As we are dealing with
small initial density contrasts defined by (\ref{small_D1}),
(\ref{small_D2}) and (\ref{small_D3}), we used for all the plots $\epsilon_1\approx 2 \rhoir/\rhoim=10^{3}$
and $\epsilon_2\approx (c^4/8\pi G)(\Ri/\rhoim)=10^{-7}$.}\label{FrhoP}
\end{figure}
\begin{figure}
\caption{The function $\Gamma$ given by (\ref{eqGamma1}) in terms of
$\log_{10}(y)$ and $\log_{10}(|\Delta_{i}^{(s)}|)$, from the initial
hypersurface $y = 1$ corresponding to $T_{i} = 10^{6}$
Kelvin. $\Gamma$ is positive as required, it is almost equals to one
for all the evolution range. (a) The lumps case, (b) the voids
case.}\label{FGamma}
\end{figure}
\begin{figure}
\caption{Plot of $\log_{10}(\tau)$ in terms of $\log_{10}(y)$ and
$\log_{10}(|\Delta_{i}^{(s)}|)$, $\tau$ is a positive definite and
increasing function for $10^{-8}\le|\Delta_{i}^{(s)}|\le10^{-3}$ and for
all evolution range.}\label{FFtau}
\end{figure}
\begin{figure}
\caption{Plot of $\log_{10}(\tau/t_{H})$ in terms of $\log_{10}(y)$
and $\log_{10}(|\Delta_{i}^{(s)}|)$; (a) for $10^{-8}\le|\Delta_{i}^{(s)} <
10^{-5.5})$ the relaxation time is not initially smaller than the
Hubble time ; (b) For the range $10^{-5.5} \le |\Delta_{i}^{(s)}| \le
10^{(-3)}$ the relaxation time is initially smaller than the Hubble
time, and it overtakes it at about $y=10^{2.4}$ corresponding to
$T_{D}=4000 K$ satisfying the required behavior.}\label{FtautH}
\end{figure}
\begin{figure}
\caption{The figure depicts the ratios
$\log_{10}(t_c/ t_{_{H}})$, $\log_{10}(\tau/t_{_{H}})$ and
$\log_{10}(t_\gamma/t_{_{H}})$
versus $\log_{10}(y)$ for $\Delta_i^{(r)}=10^{-4}$ and
$\Delta_i^{(s)}=10^{-3}$. Initially, the ratio $\tau/t_{_{H}}$ is far
larger than $t_\gamma/t_{_{H}}$ but become comparable to it near the
decoupling hypersurface defined by $t_{\gamma}=t_{_{H}}$. Note the
similarity in the qualitative behavior of these two ratios.
This plot shows also how the Compton time is important for high
temperatures (near $\log_{10}(y) = 0$) but rapidly overtakes
$t_{_{H}}$ and so Compton scattering is no longer an efficient
radiative process. Thompson scattering is very small initially but
becomes the dominant process near decoupling. The relaxation time
$\tau$ is a mesoscopic quantity that acts roughly as an "average"
timescale for all these processes.} \label{Ftautgammatc}
\end{figure}
\end{document}